\documentclass[12pt]{JHEP3}

\usepackage{amsmath}
\usepackage{amssymb}
\usepackage{bm}
\usepackage[dvips]{graphicx}
\catcode `@=11 \@addtoreset{equation}{section} \catcode `@=12
\renewcommand{\thefootnote}{\fnsymbol{footnote}}

\def\Tr{{\rm Tr}}

\title{Thermodynamics of Large $N$ Gauge Theories\\ with Chemical Potentials in a $1/D$ Expansion}

\preprint{TIFR/TH/10-09}

\author{
 Takeshi Morita\\

Department of Theoretical Physics, Tata Institute of Fundamental Research, \\
Homi Bhabha Rd, Mumbai 400005, India\\ 

\email{takeshi@theory.tifr.res.in}
 }


\abstract{
In order to understand thermodynamical properties of $N$ D-branes with chemical potentials associated with $R$-symmetry charges, we study a one dimensional large $N$ gauge theory (bosonic BFSS type model) as a first step.
This model is obtained through a dimensional reduction of a $1+D$ dimensional $SU(N)$ Yang-Mills theory and we use a $1/D$ expansion to investigate the phase structure.
We find three phases in the $\mu-T$ plane.
We also show that all the adjoint scalars condense at large $D$ and obtain a mass dynamically. 
This dynamical mass protects our model from the usual perturbative instability of massless scalars in a non-zero chemical potential.
We find that the system is at least meta-stable for arbitrary large values of the chemical potentials in $D \to \infty$ limit.
We also explore the existence of similar condensation in higher dimensional gauge theories in a high temperature limit.
In 2 and 3 dimensions, the condensation always happens as in one dimensional case.
On the other hand, if the dimension is higher than 4, there is a critical chemical potential and the condensation happens only if the chemical potentials are below it.}

\begin{document}

\renewcommand\thefootnote{\arabic{footnote}} 
\setcounter{footnote}{0}
\section{Introduction and summary}
\setlength{\baselineskip}{7mm}

Supersymmetric large $N$ gauge theories with 16 supercharges, which describe $N$ D brane dynamics, are fundamental theories in the study of the string theory.
However the thermodynamical properties of these theories with $R$-symmetry chemical potentials have not been studied sufficiently in terms of the gauge theory.
In the strong coupling regime, the gravity analysis is valid through the gauge/gravity correspondence, where the $R$-symmetry charges in the gauge theories will correspond to electric charges or angular momenta of the dual gravity systems \cite{Itzhaki:1998dd, Gubser:1998jb, Harmark:1999xt}.
Importantly, it has been shown that gravity systems with these charges have rich phase structures.
For example, in asymptotically AdS space, RN black hole/hairy black hole type transitions  were found \cite{Gubser:2008px, Nishioka:2009zj, Basu:2010uz, Bhattacharyya:2010yg} and 
their applications to condensed matter physics have been investigated \cite{Hartnoll:2008vx, Hartnoll:2009sz, Herzog:2009xv}.
Besides, in higher dimensional black holes with angular momenta, Meyers-Perry black hole/black ring type transitions happen \cite{Emparan:2008eg}.
Therefore we can expect that similar phase structures also appear in the supersymmetric gauge theories with corresponding chemical potentials.

However the gauge theories have not been studied sufficiently, since the presence of the chemical potentials makes the analysis difficult even in a weak coupling.
The reason is as follows.
Fields in the gauge theories are typically massless except in some special situations, e.g. ${\mathcal N}=4$ SYM on $R\times S^3$. (Several groups studied the chemical potentials in this case \cite{Basu:2005pj, Yamada:2006rx, Harmark:2006di, Harmark:2006ta, Harmark:2006ie, Dey:2007vt, Harmark:2007px, Hollowood:2008gp, Murata:2008bg, Elander:2008vw}.)
Then the massless scalars coupled to the chemical potentials will be unstable around the trivial vacuum and perturbative techniques do not work.
If these scalars are free, this instability is inevitable and the theory is destabilized by the chemical potentials\footnote{A naive regularization in free super Yang-Mills theory through an analytic continuation of the chemical potential to complex was proposed in \cite{Gubser:1998jb}. 
In this article, we use a different analysis and do not consider this regularization.
The imaginary chemical potential is considered in lattice gauge theory also to avoid the fermion problem in QCD \cite{deForcrand:2002ci}. }.
On the other hand, interactions involving the scalars may remove this instability.
For example, in a massless $g|\phi|^4$ theory, 
if we turn on a finite chemical potential $\mu$ for a $U(1)$ rotation,
the potential becomes $-\mu^2 |\phi|^2+g|\phi|^4$.
Although it causes the shift of the vacuum, the theory is still stable if $g>0$. 
As in this example, the understanding of the interactions in the gauge theories is essential to investigate the finite chemical potentials.

In this article, in order to consider this problem, 
we study the following one dimensional large $N$ gauge theory (a bosonic BFSS \cite{Banks:1996vh} type model),
\begin{align} 
S = \int_0^{\beta} dt \, \Tr
\left( \sum_{I=1}^{D} \frac12 \left(D_0 Y^{I}\right)^2 - \sum_{I,J}
\frac {g^2}{4} [Y^I,Y^J][Y^I, Y^J] \right).
\label{matrixqm-action}
\end{align} 
Here $Y^I$ are $SU(N)$ adjoint scalars.
The covariant derivative is defined by $D_0 Y^I =\partial_t Y^I - i [A_0, Y^I]$ and $A_0$ is an $SU(N)$ gauge field.
This model can be regarded as a dimensional reduction of a $1+D$ dimensional pure Yang-Mills theory.
The model is invariant under the $SO(D)$ rotation of $Y^I$, which is related to the $R$-symmetry in the supersymmetric gauge theory.
We will consider chemical potentials associated with $U(1)^{\lfloor D/2 \rfloor} \subset  SO(D)$.\footnote{
Here $\lfloor \cdots \rfloor$ denotes the floor function which maps a number to the integer part.}
Note that the charges of these $U(1)^{\lfloor D/2 \rfloor}$ correspond to the transverse angular momenta on $\lfloor D/2 \rfloor$ planes in the context of the rotating D-branes \cite{Gubser:1998jb, Harmark:1999xt,  Cvetic:1996ek, Csaki:1998cb, Hawking:1998kw, Kraus:1998hv, Cvetic:1999ne, Cvetic:1999rb, Hawking:1999dp}.

We can obtain the model (\ref{matrixqm-action}) with $D=9$ from $N$ D1-branes on a spatial circle as follows \cite{Aharony:2005ew}.
We can choose the boundary conditions of the fermions for this circle as anti-periodic.
Then all the fermions obtain a mass, which is proportional to the inverse radius of the circle.
Thus, if the radius is sufficiently small, we can ignore all the fermions, KK-modes and long string states and the theory is reduced to one dimension.
Then the $SO(D-1)$ symmetry is enhanced to $SO(D)$ and we obtain (\ref{matrixqm-action}).
Note that it is known that the gauge/gravity correspondence is not valid in such a small radius, since the effective coupling becomes small \cite{Aharony:2005ew}. 
Thus we cannot compare our results with the gravity directly.
Our results should be regarded as a weak coupling continuation of the gravity results. 
(However, as far as I know, there is no known gravity result of the thermodynamics of this system with the chemical potentials.)

The thermodynamics of the model (\ref{matrixqm-action}) without the chemical potential has been investigated by using a $1/D$ expansion \cite{Mandal:2009vz}.
Especially it has been revealed that a non-trivial stable vacuum exists in which the adjoint scalars condense and obtain a mass. 
We generalize this analysis to the finite chemical potentials. 
We will show that, in the $D \to \infty$ limit, the (meta-)stable condensed vacuum always exists for arbitrary values of the chemical potentials.
This result is similar to the $g|\phi|^4$ theory and we can conclude that the commutator-squared interaction stabilizes our model in the large $D$ limit.
However, in a large but finite $D$ case, the $1/D$ expansion does not converge for large chemical potentials and it is unclear whether the system is stable or not there.

We investigate the phase structure also and show the existence of three phases in the $\mu-T$ plane, which is summarized in Figure \ref{fig phase diagram}.

In a single chemical potential case, we find several saddle points in a high temperature regime.
In this regime, since only the zero-modes of the Matsubara frequencies are dominant, 
the model (\ref{matrixqm-action}) will reduce to an effective zero-dimensional matrix model (bosonic IKKT type model \cite{Ishibashi:1996xs}).
 In this model, the chemical potential induces a CS like interaction.
(However, its coefficient is real and different from the ordinary CS term in \cite{Iso:2001mg}.)
Therefore, (imaginary) fuzzy sphere like solutions exist as complex saddle points.
However, physical interpretation of these solutions is yet to be explored.\\

We also explore the existence of the condensed vacuum in the $d$ dimensional large $N$ gauge theory to understand the stability issue.\footnote{
The $1/D$ expansion analysis of $d$ dimensional gauge theory without the chemical potentials for $d=0$ (bosonic IKKT model) is considered in \cite{Hotta:1998en} and
for $d=2$ (two-dimensional gauge theory on $T^2$) is considered in an ongoing work  \cite{Dynamical}.
}
Since the analysis is difficult in general, we restrict our study to a high temperature regime and consider the $D \to \infty$ limit.
Then we will see that the existence of the condensation depends on the dimension $d$.
In $d=2,3$ cases, the condensation always happens for arbitrary values of the chemical potentials as in the $d=1$ case.
On the other hand, in $d \ge 4$ case, a critical chemical potential $\mu_{c}$ exists and the condensation happens only if  all the chemical potentials are below $\mu_{c}$.
These results are summarized in Figure \ref{fig large d phase}.

The same condensations will happen in the supersymmetric gauge theories also, since we can ignore the fermions in the high temperature limit.
Thus our results may be related to the analysis in the strong coupling regime through the gravitational study of the rotating D brane geometries \cite{Harmark:1999xt}.
However the application of our large $D$ results to the supersymmetric gauge theories are not robust particularly in a high chemical potential regime and further analyses are necessary.\\

The organization of this article is as follows.
In section \ref{sec 1dMM}, we introduce the chemical potentials to our model (\ref{matrixqm-action}) and, by using the large $D$ expansion, we derive an effective action.
We will analyze this effective action in two regimes.
One is a low temperature and low chemical potential regime.
Another is a high temperature regime and/or a high chemical potential regime.
We will study the phase structure of the first regime in section \ref{Section low T}.
In section \ref{Section high T}, we investigate the nature of the second regime.
We will also discuss the fuzzy sphere like saddle points in this section.
In section \ref{sec high d}, we explore the existence of the condensation in higher dimensional gauge theories.
In section \ref{Conclusion}, we conclude with a discussion.

\section{One-dimensional gauge theory with chemical potential}
\label{sec 1dMM}

\subsection{Chemical potential}
\label{sec chemical}
We consider the large $N$ gauge theory (\ref{matrixqm-action}) with chemical potentials associated with the global $U(1)^{\lfloor D/2 \rfloor}$ transformations.
It is convenient to rename the $D$ adjoint scalar $Y^I$ as
\begin{align} 
X^I& \equiv Y^I~(I=1,\dots,\lfloor D/2 \rfloor), \nonumber \\
W^I &\equiv Y^I~(I=\lfloor D/2 \rfloor+1,\dots,2 \lfloor D/2 \rfloor), \nonumber \\
Y^D &= Y^D~(\text{If $D$ is even, this one does not exist.}),
\end{align} 
and define complex scalars as $\Phi^I \equiv (X^I+iW^I)/\sqrt{2}$, ($I=1,\dots,\lfloor D/2 \rfloor$).
Then the action (\ref{matrixqm-action}) is invariant under the global $U(1)^{\lfloor D/2 \rfloor}$ transformation: $\Phi^I \to e^{i\Lambda_I} \Phi^I$.
In the context of the rotating D-brane, this charge corresponds to the angular momentum on the $(X^I, W^I)$ plane.

Now we introduce the chemical potential $\mu_I$ for these $U(1)$ transformations.
It has been studied in \cite{Yamada:2006rx, Haber:1981ts} that, in the Euclidean path integral, the chemical potential causes the modification of the action as
\begin{align} 
D_0 \to D_0 - \mu_I  \quad (I=1,\ldots, \lfloor D/2 \rfloor),
\label{chemical potential}
\end{align} 
in the kinetic term of $\Phi^I$.
Without loss of generality, we can take $\mu_I$ positive.
This modification gives rise to negative mass term $-\mu_I^2 |\Phi^I|^2$ and the system might be destabilized. In addition, the usual perturbative analysis does not work.

In the following sections, we will investigate the thermodynamics of the gauge theory by using a $1/D$ expansion \cite{Mandal:2009vz, Hotta:1998en}, in which the number of the adjoint scalars is regarded as large.
In order to keep the contribution of the chemical potentials finite under the large $D$ limit, 
first we consider a simple situation,
\begin{align} 
 \mu_I&=\mu \qquad ( I=1,\cdots, \tilde{D} ), \nonumber \\
\mu_i&=0     \qquad  (i=\tilde{D}+1, \cdots, \lfloor D/2 \rfloor),  
\label{simple mu}
\end{align} 
and $\tilde{D}$ scales linearly with $D$.
If we consider general chemical potential $\mu_I$, each contribution will appear as sub-leading in the $1/D$ expansion. 
Thus in order to evaluate them consistently, we have to derive other sub-leading terms also and compare with them. 
We will study it in section \ref{sec general mu} and section \ref{sec D=1}.

The validity of the application of the $1/D$ expansion to D1-brane case ($D=9$) is not a priori obvious, since $D$ is large but finite.
However in the case of the zero-chemical potential, the $1/D$ expansion \cite{Mandal:2009vz} reproduces the numerical results in \cite{Aharony:2005ew, Aharony:2004ig,  Kawahara:2007fn, Azuma:2007fj, Azeyanagi:2009zf} quantitatively.
In fact the leading large $D$ results even for $D=2,3$ reproduce the essential qualitative properties. 
Thus we expect that the $1/D$ expansion is also meaningful in the model involving the  chemical potentials.

\subsection{$1/D$ expansion and computation of effective action}
\label{sec Eff}
In this subsection, we will integrate out the adjoint scalars in the action (\ref{matrixqm-action}) with the chemical potential (\ref{chemical potential}) and derive an effective action by using the $1/D$ expansion.
We evaluate only the leading order of this expansion in this section.
We will discuss the contribution from the next order in section \ref{sec 1/D} and \ref{sec high mu 1/D }.
In section \ref{sec 1/D}, it will turn out that inclusion of these corrections does not change the nature of the phase structure in a low chemical potential regime.
On the other hand, in section \ref{sec high mu 1/D }, we will show that the $1/D$ expansion does not converge in a very high chemical potential regime.

According to \cite{Mandal:2009vz}, we rewrite the path integral by employing an auxiliary field $B_{ab}$ as\footnote{The definition of $B_{ab}$ is different from  \cite{Mandal:2009vz} by an imaginary factor ``$i$".}
\begin{align} 
Z= & {\cal N} 
\int {\cal D} B {\cal D} A_0 {\cal D} Y^i {\cal D} \Phi^I e^{-S(B,A_0,Y,\Phi)}, \nonumber \\
S(B,A_0,Y,\Phi) =& \int_0^\beta dt\, \Biggl[
- \frac{1}{4g^2} B_{ab} M^{-1}_{ab,cd}B_{cd} \nonumber \\
 & +\sum_{I=1}^{\tilde{D}} \Phi^{\dagger I}_a \left(-(D_0-\mu)^2_{ab}+ B_{ab} \right)  \Phi^{I}_b
+\sum_{i=2\tilde{D}+1}^D \frac12  Y^{i}_a \left( - D_{0ab}^2+B_{ab} \right) Y^{i}_b  
 \Biggr]. 
\label{gauss-trick}
\end{align} 
Again we have used the notation $Y^i$ for the adjoint scalars, which do not couple to the non-zero chemical potentials.
Here $1/{\cal N} \equiv \int {\cal D} B \exp\left( \int dt  B_{ab} M^{-1}_{ab,cd}B_{cd}/(4g^2) \right)  $ is a numerical factor.
$M_{ab,cd}^{-1}$ is the inverse of $M_{ab,cd}$, which is defined by
\begin{align} 
M_{ab,cd} = -\frac{1}{4} \Bigl\{ \Tr[\lambda_a,
  \lambda_c][\lambda_b, \lambda_d] +(a\leftrightarrow b)+(c\leftrightarrow
d)+(a\leftrightarrow b,c\leftrightarrow d) \Bigr\}.
\label{def Mabcd}
\end{align}
Here $\lambda_a$ $(a=1\dots N^2-1)$ is a generator of $SU(N)$.
They satisfy $M_{ab,cd} M_{cd,ef}^{-1}=(\delta_{ae}\delta_{bf}+\delta_{af}\delta_{be})/2$. Properties of the matrix $M_{ab,cd}$ is summarized in appendix A of \cite{Mandal:2009vz}.
We can reproduce the original action (\ref{matrixqm-action}) from (\ref{gauss-trick}) by substituting the solution of the classical equation of motion for $B_{ab}$:
\begin{align} 
M^{-1}_{ab,cd}B_{cd}=g^2\left(\sum_{I=1}^{\tilde{D}} \left(  \Phi^{I\dagger}_a\Phi^{I}_b +  \Phi^{I\dagger}_b\Phi^{I}_a \right) + \sum_{i=2\tilde{D}+1}^D  Y_a^iY_b^i  \right) .
\label{classical sol B}
\end{align} 

Now we can formally integrate out $\Phi^I$ and $Y^i$,
since the action (\ref{gauss-trick}) is quadratic in them.
Then we obtain an effective action for $B_{ab}$ and $A_0$,
\begin{align} 
S_{eff}(B,A_0) =& \int_0^\beta dt\, \Biggl[
- \frac{1}{4g^2} B_{ab} M^{-1}_{ab,cd}B_{cd} \Biggr] \nonumber \\
 &+\frac{k_1 D}{2} \log \det \left( - D_{0ab}^2+B_{ab} \right)  
 +\frac{k_2 D}{2}  \log \det \left(-(D_0-\mu)^2_{ab}+ B_{ab} \right)  ,
\label{gauss-trick-2}
\end{align} 
where we have defined normalized ratios: $k_1=(D-2\tilde{D})/D$ and $k_2=2\tilde{D}/D$, which satisfy $k_1+k_2=1$.
As in \cite{Mandal:2009vz},  we investigate this model by taking the large $D$ and large $N$ limit such that $D \to \infty$, $N \to \infty$ and $g \to 0$ with fixed $\tilde{\lambda}\equiv g^2 N D =\lambda D$.
(In our case, we also take $\tilde{D} \to \infty$ with fixed $k_2=2\tilde{D}/D$).
Then the first term and the $\log \det$ terms in (\ref{gauss-trick-2}) will be comparable in this limit.
As a result, this model will have a non-trivial saddle point $\bar{B}_{ab}=\triangle_0^2 \delta_{ab}$, which we will confirm later\footnote{The large $N$ limit is not necessary to derive this saddle point. Actually, even at finite $N$, we can calculate some physical quantities by taking $D \to \infty$ and $g^2N \to 0$ with fixed $g^2N D$ \cite{Mandal:2009vz, Hotta:1998en}. However our interest is in the large $N$ limit of (\ref{matrixqm-action}) and we do not consider finite $N$ effects in this article.}. 
Here $\triangle_0$ is a time independent constant.
From equation (\ref{classical sol B}), indeed this saddle point gives a condensation of the adjoint scalars
\begin{align} 
2 \sum_{I=1}^{\tilde{D}} \langle \Tr \Phi^{I\dagger}\Phi^{I} \rangle  + \sum_{i=2\tilde{D}+1}^D  \langle \Tr  Y^iY^i   \rangle
= \frac{N}{2g^2}\triangle_0^2 ,
\end{align} 
where we have used a relation $M_{ab,cd}^{-1} \delta_{cd}=\frac{1}{2N}\delta_{ab}$  \cite{Mandal:2009vz}.

To proceed, we write the $B_{ab}$ as the sum of a constant trace piece and the rest as
\begin{align}
B_{ab}(t) = \triangle^2 \delta_{ab} +g b_{ab}(t),
\label{fluct}
\end{align}
where $b_{ab}(t)$ satisfies $\int dt\, b_{aa}(t)=0$.
Note that we can ignore the interactions between $b_{ab}$ and $\Phi^I$ and $Y^i$ in the leading order of the $1/D$ expansion \cite{Mandal:2009vz}.
The contributions of the interactions will appear in the next order as we will see in section \ref{sec 1/D} and \ref{sec high mu 1/D }.

Here we consider a gauge fixing of $A_0$.
It is convenient to take the constant diagonal gauge: $A_{0ij}(t)=\alpha_{i}\delta_{ij}$.
As a result the effective action will be described by gauge invariant Wilson loop operators,
\begin{align}
u_n = \frac{1}{N} \Tr  e^{in \int_0^\beta A_0 dt}= \frac{1}{N} \sum_{i=1}^N e^{in\beta \alpha_i}.
\end{align} 
This gauge fixing gives rise to a Faddeev Popov determinant \cite{Aharony:2003sx}
\begin{align} 
{\cal D}A_0=\prod_i d\alpha_i e^{-S_{FP}},\quad S_{FP}=N^2\sum_n 
\frac1{n} |u_n|^2.
\label{FP}
\end{align} 

Now we integrate out $b_{ab}$ and obtain the effective action for the condensation $\triangle$ and the gauge field $\{u_n \}$.
\begin{align} 
\frac{S_{eff}(\triangle,\{u_n\})}{DN^2}=& -\frac{\beta \triangle^4}{8 \tilde{\lambda}}+  \frac{1}{D} 
\sum_{n=1}^\infty 
\frac{|u_n|^2}{n} \nonumber \\
&+ \frac{1}{2N^2}  \left[ k_1 \Tr \log \left(-D_0^2 + \triangle^2 \right) 
+ k_2 \Tr \log \left(-\left( D_0-\mu \right)^2 + \triangle^2 \right) 
\right].
\label{Seff process}
\end{align} 
Here the first term is a classical term from the first term in (\ref{gauss-trick-2}).
The second term comes from (\ref{FP}) and indeed it is $1/D$ order.
We keep it here, since this term will be dominant in a low temperature regime and more significant than other $O(1/D)$ terms.
The integral of $b_{ab}$ gives a numerical factor which cancels out ${\mathcal N}$ in (\ref{gauss-trick}) except the contribution of the integral of the constant trace peace $\triangle$.

Let us evaluate the terms in the second line of (\ref{Seff process}).
In the momentum space, the quadratic term of $\Phi^I$ in (\ref{gauss-trick}) can be written as
\begin{align} 
\frac{1}{2} 
\begin{pmatrix} 
X_{nij}^I &W_{nij}^I \end{pmatrix} 
\begin{pmatrix} 
\left(\frac{2\pi n}{\beta} -(\alpha_j-\alpha_i) \right)^2 -\mu^2 + \triangle^2& -2\mu \left(\frac{2\pi n}{\beta} -(\alpha_j-\alpha_i) \right) \\
2\mu \left(\frac{2\pi n}{\beta} -(\alpha_j-\alpha_i) \right)&\left(\frac{2\pi n}{\beta} -(\alpha_j-\alpha_i) \right)^2 -\mu^2 + \triangle^2\end{pmatrix} 
\begin{pmatrix} 
X^I_{-nji} \\ W_{-nji}^I \end{pmatrix} .
\label{phi kinetic}
\end{align} 
Then the eigenvalues of this matrix are calculated as
\begin{align} 
 \left(\frac{2\pi n}{\beta} -(\alpha_j-\alpha_i)  \pm i \mu \right)^2  + \triangle^2.
\end{align} 
By using this result, we calculate \cite{Yamada:2006rx, Aharony:2003sx}
\begin{align} 
&\Tr \log \left(-\left( D_0-\mu \right)^2 + \triangle^2 \right) \nonumber \\
=&\frac{1}{2} \sum_{n,i,j}
\left[
 \log \left(  \left(\frac{2\pi n}{\beta} -(\alpha_j-\alpha_i)  + i \mu \right)^2  + \triangle^2
 \right)+\log \left(  \left(\frac{2\pi n}{\beta} -(\alpha_j-\alpha_i)  - i \mu \right)^2  + \triangle^2
 \right) \right] \nonumber \\
 =&\sum_{n,i,j} \log \left[  \tilde{{\mathcal N}}e^{\beta \triangle} 
\left( 1- e^{-\beta \triangle+i\beta \left( \alpha_i-\alpha_j \right)-\beta \mu } \right) 
\left( 1- e^{-\beta \triangle-i\beta \left( \alpha_i-\alpha_j \right)+\beta \mu } \right) \right] \nonumber \\
 =& N^2 \log \tilde{{\mathcal N}}+ N^2 \beta \triangle 
-N^2 \sum_{n=1}^\infty \frac{1}{n}  e^{-n\beta \triangle }
\left( e^{n\beta \mu }+e^{-n\beta \mu }\right) 
|u_n|^2,
\label{log det}
\end{align} 
where we have ignored $O\left( 1/N^2\right) $ corrections.
$\tilde{{\mathcal N}}$ is a irrelevant constant factor and we will ignore it from now.
Similarly we evaluate the term from $Y^I$ integral also.
Then the effective action (\ref{Seff process}) becomes
\begin{align} 
\frac{S_{eff}(\triangle,\{u_n\})}{DN^2}=& -\frac{\beta \triangle^4}{8 \tilde{\lambda}}+  \frac{1}{D} 
\sum_{n=1}^\infty 
\frac{|u_n|^2}{n} \nonumber \\
&+\frac{\beta \triangle}{2} -\sum_{n=1}^\infty 
\left[ e^{-n\beta \triangle}
\left(k_1+k_2 \frac{e^{-n\beta \mu}+e^{n\beta \mu}}{2}\right)
\right] 
\frac{|u_n|^2}{n} .
\label{potential triangle}
\end{align} 
Note that, during this derivation, we have assumed a relation $\triangle \ge \mu$.
If this inequality is not satisfied, tachyonic modes will appear and the path integral is not well defined\footnote{$\triangle=\mu$ case is also subtle, since massless modes will appear and the effective action will be non-local. Indeed it will happen at $T=0$. We will come back to this problem later.}.
Later we will show that this assumption is ensured.

From now we evaluate the effective action (\ref{potential triangle}) and investigate the phase structure and the condensation of the scalars.
It is convenient to integrate out $\triangle$ first and derive the effective action for the Wilson loops.
In order to do it, we analyze the saddle point equation for $\triangle^2$, \footnote{
When we derive the saddle point equation (\ref{saddle point triangle}), we deviate the effective action with respect to $\triangle^2$ (not $\triangle$), since $\triangle^2$ is the correct variable as in (\ref{fluct}).}
\begin{align} 
-\frac{\triangle^3}{2\tilde{\lambda} } +\frac{1}{2} +\sum_{n=1}^\infty 
e^{-n\beta \triangle}
\left[ k_1+k_2
\left(
 \frac{e^{-n\beta \mu}+e^{n\beta \mu}}{2} \right)
\right] 
|u_n|^2=0.
\label{saddle point triangle}
\end{align} 
We will solve this equation in two different regimes characterized by the values of the Wilson loop operators. 
One is the low temperature and low chemical potential regime ($|u_n| \sim 0$, $n\ge 2$) and another is the high temperature or high chemical potential regime ($|u_n|\sim 1$).
We will study the first regime in section \ref{Section low T} and the second one in section \ref{Section high T}.

Before proceeding the analysis of our model, we remark about the validity of the $1/D$ expansion.
In the strict large $D$ limit, 
our analysis will be valid for any temperature and chemical potentials.
However, in a large but finite $D$ case, some problems about the $1/D$ expansion will arise in a very low temperature regime and a very high chemical potential regime.
In the very low temperature regime, the dimensionless effective coupling $\tilde{\lambda}/T^3$ will be large. 
Thus the contributions of higher loops will be large and the expansion will not work.
In the very high chemical potential regime, as we will see later, light mass modes will appear and similarly the contribution of the higher loops will become large.
We will discuss the details of these issues in section \ref{sec 1/D} and \ref{sec high mu 1/D }.

We summarize the regimes which we will consider in this article in Table \ref{table regime}. 
\TABLE{
\begin{tabular}{lc}
\hline
very low temperature regime & \lower .3ex\hbox{$T/\tilde{\lambda}^{1/3}< D^{-\gamma}$} \\
low temperature and low chemical potential regime & $|u_n| \sim 0$, $n\ge 2$ \\
intermediate temperature and chemical potential regime & $|u_n| \ne 0$, $n\ge 2$ \\
high temperature regime & $|u_n| \sim 1$ \\
high chemical potential regime & $|u_n| \sim 1$ \\
very high chemical potential regime & $\mu/\tilde{\lambda}^{1/3} \gg (k_2 T/\tilde{\lambda}^{1/3})^{1/4} $ \\
\hline
\end{tabular}
\caption{Various regimes in the analysis of the one dimensional gauge theory.
$T/\tilde{\lambda}^{1/3}$ and $\mu/\tilde{\lambda}^{1/3}$ are dimensionless temperature and chemical potential.
We will discuss our model in the first three regimes in section \ref{Section low T} and the rest in section \ref{Section high T}.
}
\label{table regime}
}

\section{The phase structure of the low temperature and low chemical potential regime}
\label{Section low T}

\subsection{The phase structure in the leading $1/D$ expansion}
\label{sec low T}

In this subsection, we evaluate the phase structure in the low temperature and low chemical potential regime, by analyzing the effective action (\ref{potential triangle}).
Note that this effective action is the leading order of the $1/D$ expansion.
We will consider next order corrections in subsection \ref{sec 1/D} but, as we have mentioned, they do not change the nature of the phase structure.
The obtained phase structure is summarized in Figure \ref{fig phase diagram}.

\FIGURE{
\includegraphics[scale=0.75]{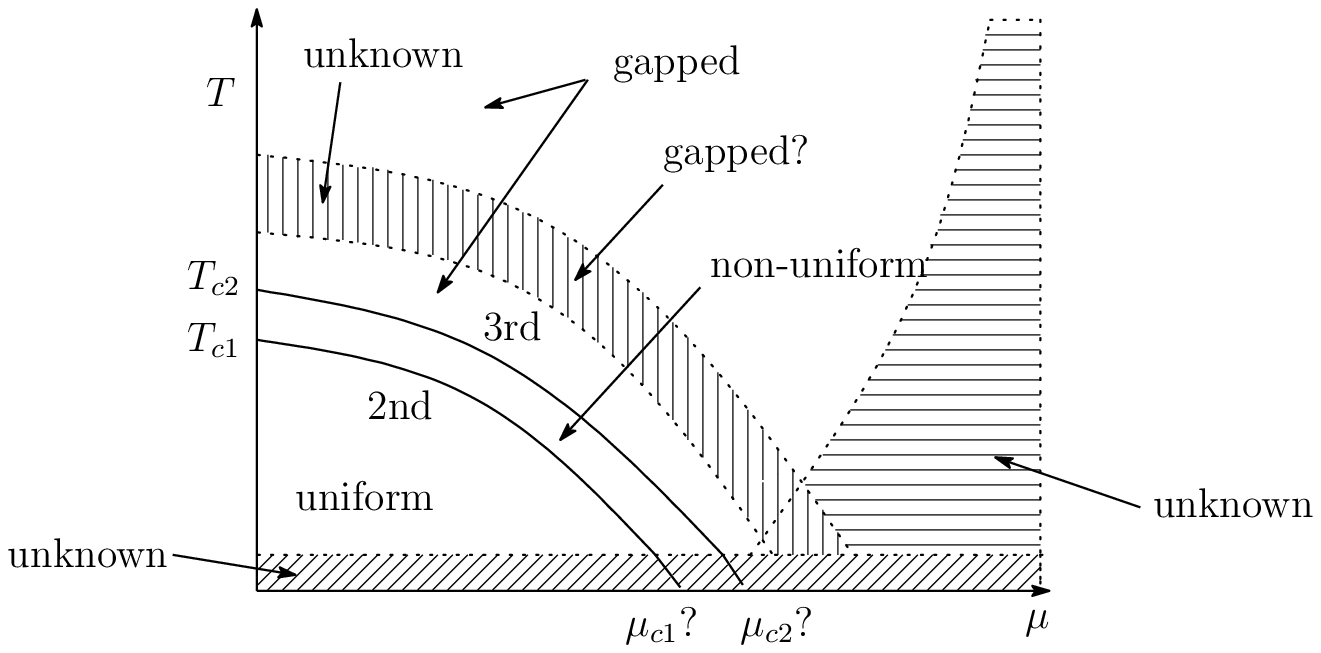}
\caption{Phase diagram of the one dimensional gauge theory in $\mu-T$ space from the $1/D$ expansion.
Three (uniform, non-uniform and gapped) phases exist and the orders of the phase transitions between them are second and third. 
In the shaded regions, it is difficult to analyze the model through the $1/D$ expansion.
In the horizontal shaded region (very high chemical potential region $\mu/\tilde{\lambda}^{1/3} \gg (k_2 T/\tilde{\lambda}^{1/3})^{1/4} $), the expansion does not converge because of the existence of the light mass modes.
In the inclined shaded region (very low temperature region $T/\tilde{\lambda}^{1/3} < D^{-\gamma}$), the expansion is not valid, since the effective coupling $\tilde{\lambda}/T^3$ becomes too strong.
The analysis in the vertically shaded region is also difficult, since the Wilson loop operators are highly interacting (intermediate region).
However we can guess that the vertically shaded region will be gapped phase.  
See Table \ref{table regime} also.}
\label{fig phase diagram}}

If both of the temperature and chemical potential are sufficiently low such that all the coefficients of $|u_n|^2$ in (\ref{potential triangle}) are positive, the stable solution is given by $|u_n|=0$ for all $n$.
Thus the contributions of $|u_n|$ are small in this regime and it is indeed enough to keep only $|u_1|$ in the saddle point equation (\ref{saddle point triangle}) to analyze the thermodynamics in this regime,
\begin{align} 
\frac{\triangle^3}{\tilde{\lambda} } =1 +
2e^{-\beta \triangle}
\left( k_1+
k_2\left(
\frac{ e^{-\beta \mu}+e^{\beta \mu}  }{2}\right)\right)
|u_1|^2.
\label{saddle point low T}
\end{align} 
Since $e^{-\beta \triangle}$, $e^{-\beta (\triangle-\mu)}$ and $e^{-\beta (\triangle+\mu)}$ will be small in this regime, we can solve this equation approximately and obtain the condensation as
\begin{align} 
\frac{\triangle}{\tilde{\lambda}^{1/3}}=1+ \frac{2}{3} 
e^{-\beta \tilde{\lambda}^{1/3} }
\left( k_1+k_2
\left(
\frac{
 e^{-\beta \mu}+e^{\beta \mu}}{2}
\right)\right)
|u_1|^2+\cdots.
\label{triangle low T}
\end{align} 
Then, by putting this solution into (\ref{potential triangle}), we obtain an effective action for the Wilson loops
\begin{align} 
\frac{S_{eff}(\{u_n\})}{DN^2}=& \frac{3\beta \tilde{\lambda}^{1/3} }{8}  
+  a(\beta,\mu)  |u_1|^2+ b(\beta,\mu)  |u_1|^4+  \frac{1}{D} 
\sum_{n=2}^\infty 
\frac{|u_n|^2}{n}+\cdots,
\label{action wilson loop}
\end{align} 
where the coefficient $a(\beta,\mu)$ and $b(\beta,\mu)$ are given by
\begin{align} 
a(\beta,\mu)&=  \frac{1}{D} 
-e^{-\beta \tilde{\lambda}^{1/3}}
\left( k_1+
k_2\left(
 \frac{e^{-\beta \mu}+e^{\beta \mu}}{2} \right) \right),
\label{solution a}
 \\
b(\beta,\mu)&=\frac{\beta\tilde{\lambda}^{1/3}}{3}e^{-2\beta \tilde{\lambda}^{1/3}} 
\left( 
k_1+k_2\left(
 \frac{e^{-\beta \mu}+e^{\beta \mu}}{2} \right)  \right)^2.  
\end{align} 
Note that $b(\beta,\mu)$ is always positive.
In this case, three phases will appear as we will show soon through the argument of the Landau-Ginzburg type analysis in \cite{Aharony:2003sx, AlvarezGaume:2005fv}.
(If $b<0$, two phases will appear instead of them.)
Here the order parameters of these phases will be the values of the Wilson loop operator $u_n$ or equivalently the eigenvalue density of the gauge field $A_0$, which is defined as
\begin{align} 
\rho(\alpha)\equiv& \frac{1}{N}\sum_{i=1}^N \delta(\alpha-\alpha_i) = \frac{\beta}{2\pi}\left(1+\sum_{n\ne 0} u_n e^{-i n\beta  \alpha} \right)  .
\end{align} 

\FIGURE{
\includegraphics[scale=0.75]{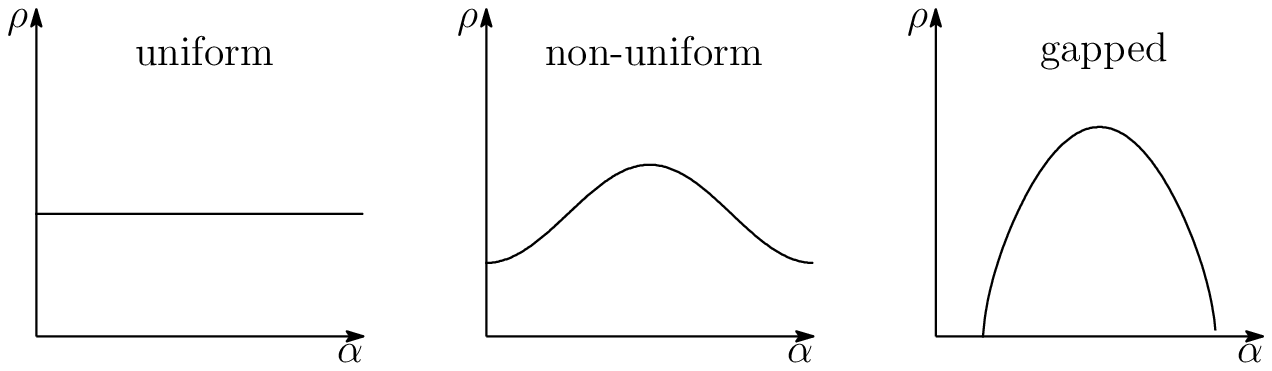}
\caption{Plots of eigenvalue density function $\rho(\alpha)$.
Three configurations of $\rho(\alpha)$ characterize the three phases in the $\mu-T$ phase diagram.
}
\label{fig rho}
}

Now we investigate the three phases. 
If the temperature and chemical potential are both low, $a(\beta,\mu)>0$ is satisfied.
Then, from the effective action (\ref{action wilson loop}), the stable configuration is $|u_n|=0$ for all $n$ as we have mentioned.
This phase is called uniform phase, since the eigenvalue density $\rho(\alpha)$ is constant and thus it is uniform with respect to $\alpha$.
(See Figure \ref{fig rho}.)
This phase is an analogue of the confinement phase in the higher dimensional gauge theory,
 since the expectation values of the temporal Wilson loops vanish.

As the temperature or chemical potential increases, $a(\beta,\mu)$ becomes 0.
On the curve $a(\beta,\mu)=0$, still $|u_n|=0$ is stable since $b(\beta,\mu)$ is positive.
However if $a(\beta,\mu)<0$, the solution $|u_1|=0$ becomes unstable and a stable solution, which is given by $|u_1|=\sqrt{-a/2b}$ and $|u_n|=0$ for $n \ge 2$, appears.
The configuration of the eigenvalue density $\rho(\alpha)$ becomes non-uniform as shown in Figure \ref{fig rho} and this phase is called non-uniform phase.
Thus a phase transition happens on $a(\beta,\mu)=0$.
This line is the first phase transition line in $\mu-T$ phase space. (See Figure \ref{fig phase diagram}.) 
It is easy to show that this transition is second order by evaluating the free energy \cite{Mandal:2009vz, Aharony:2003sx}.

As the temperature or chemical potential increases further, $|u_1|$ achieves 1/2.
Then a gap appears in the eigenvalue density $\rho$. 
(We can choose $u_1$ real by a gauge fixing, and then the gap arises at $\alpha=\pm \pi/\beta.$ )
This is a Gross-Witten-Wadia type third order phase transition \cite{Gross:1980he,
Wadia:1980cp}.
In this phase, all the Wilson loop operators become non-zero.
This phase is called gapped phase and is an analogue of the deconfinement phase in the higher dimensional gauge theory.
The curve $|u_1|=\sqrt{-a/2b}=1/2$ gives the second phase transition line in Figure \ref{fig phase diagram}.

We have found the two phase transition lines between the three phases.
Let us analyze the details of these two curves.
First we evaluate the curve described by $a(\beta,\mu)=0$.
For small $\mu$,  we can solve this equation and obtain $T$ as a function of $\mu$,\footnote{Since the mass dimension of $\tilde{\lambda}$ is 3 in our model, $T/\tilde{\lambda}^{1/3}$ and $\mu/\tilde{\lambda}^{1/3}$ can be regarded as dimensionless temperature and chemical potential.}
\footnote{
The critical temperature (\ref{Tc1}) goes to zero if we take $D \to \infty$.
Similarly the second critical temperature (\ref{Tc2}) will also go to zero in this limit.
(In this limit, even though the critical temperatures become very low, the $1/D$ corrections are suppressed further and these results are exact.)
As a result, only the gapped phase appears in $D=\infty$.
 See Figure \ref{fig large d phase}.}
\begin{align} 
\frac{T_{c1}(\mu)}{\tilde{\lambda}^{1/3}} =\frac{1}{\log D} - \frac{k_2}{2}\left(\frac{\mu}{\tilde{\lambda}^{1/3} }  \right)^2+\cdots.  
\label{Tc1}
\end{align} 
Here we can see that the existence of the chemical potential reduces the critical temperature.
It means the chemical potential enhances the non-uniform phase.
Note that, at $\mu=0$, this curve is coincident with the result in \cite{Mandal:2009vz}.

For finite $\mu$,
the $e^{\beta \mu}$ term in (\ref{solution a}) is dominant and 
the curve is described as
\begin{align} 
\frac{T_{c1}(\mu)}{\tilde{\lambda}^{1/3}}=
\frac{1}{ \log \tilde{D}}\left(1-\mu/\tilde{\lambda}^{1/3} \right) 
+\cdots.
\label{eq mu-c1}
\end{align} 
Thus as $\mu/\tilde{\lambda}^{1/3}$ approaches 1, the critical temperature goes to 0, and $\mu_{c1}/\tilde{\lambda}^{1/3}=1$ seems a critical chemical potential at $T=0$.
However the $1/D$ expansion will not be valid in such a very low temperature regime and this result is not reliable.
Indeed some strange things will happen around this point when we consider $1/D$ corrections in subsection \ref{sec 1/D}.

Next we evaluate the second phase transition line $|u_1|=\sqrt{-a/2b}=1/2$.
For small $\mu$, the curve is given by
\begin{align} 
\frac{T_{c2}(\mu)}{\tilde{\lambda}^{1/3}}
=
\frac{T_{c1}(\mu)}{\tilde{\lambda}^{1/3}}
\left( 1+\frac{2}{3D} 
\left(1+k_2(\log D)^2
\left(\frac{\mu}{\tilde{\lambda}^{1/3} } \right)^2
\right)\right)  +\cdots.
\label{Tc2}
\end{align} 
For finite $\mu/\tilde{\lambda}^{1/3}$ $(< 1 )$ ,
the curve behaves
\begin{align} 
\frac{T_{c2}(\mu)}{\tilde{\lambda}^{1/3}} =\frac{T_{c1}(\mu)}{\tilde{\lambda}^{1/3}}\left(1+\frac{2}{3D}\frac{1}{1-\mu/\tilde{\lambda}^{1/3} }   \right) +\cdots.
\end{align} 
However this equation is not valid around $\mu/\tilde{\lambda}^{1/3} \sim 1 $.
In order to investigate this region, it is convenient to evaluate the saddle point equation for $u_1$, which is derived from (\ref{potential triangle}), as
\begin{align} 
\left[
\frac{1}{D}- 
 e^{-\beta \triangle}
\left(k_1+k_2 \frac{e^{-\beta \mu}+e^{\beta \mu}}{2}\right)
\right] u_1=0.
\end{align} 
Thus in case $u_1 \ne 0$, $\triangle$ has to satisfy
\begin{align} 
 e^{-\beta \triangle}
\left(k_1+k_2 \frac{e^{-\beta \mu}+e^{\beta \mu}}{2}\right)
=\frac{1}{D}.
\label{triangle non-uniform 1}
\end{align} 
Since $\beta$ will be large on the curve near $\mu/\tilde{\lambda}^{1/3}  \sim 1$,
we can approximately solve this equation as
\begin{align} 
\triangle=\mu+\frac{1}{\beta}  \log \tilde{D}+\cdots
\label{triangle non-uniform 2}.
\end{align} 
By putting it into the saddle point equation (\ref{saddle point low T}), we obtain
\begin{align} 
|u_1|^2=\frac{D}{2}\left(\frac{1}{\tilde{\lambda}}
\left(\mu+\frac{1}{\beta}  \log \tilde{D} \right)^3-1     \right)  +\cdots.
\end{align} 
The positivity of $|u_1|$ requires that this solution is valid only if $\triangle^3/\tilde{\lambda}^{1/3} \ge 1 $.
Now we can derive the second phase transition line around $\mu/\tilde{\lambda}^{1/3} \sim 1 $ by putting $|u_1|=1/2$ in this equation,
\begin{align} 
\frac{T_{c2}(\mu)}{\tilde{\lambda}^{1/3}} =1-\mu/\tilde{\lambda}^{1/3} +\frac{1}{6\tilde{D} } +\cdots.
\label{mu-c2}
\end{align} 
Although this equation predicts a critical value of the chemical potential $\mu_{c2}/\tilde{\lambda}^{1/3}=1+1/6\tilde{D}$ at $T=0$, the $1/D$ expansion will not work there.

Finally let us confirm that the relation $\triangle \ge \mu$, which we have assumed, is always satisfied in the uniform and non-uniform phase.
In the uniform phase, $\triangle/\tilde{\lambda}^{1/3}=1$ from (\ref{triangle low T}).
Since the uniform phase exists up to $\mu_{c1}/\tilde{\lambda}^{1/3} = 1$, the relation  $\triangle \ge \mu$ is satisfied.
In the non-uniform phase, the equation (\ref{triangle non-uniform 1}) and (\ref{triangle non-uniform 2}) show $\triangle \ge \mu$.

A problem is the case $\triangle=\mu$, which arises on a line $T=0$, $\mu/\tilde{\lambda}^{1/3} \ge 1 $.
It causes zero modes of the adjoint scalar $\Phi^I$ in (\ref{phi kinetic}).
In addition, the $1/D$ expansion itself is not valid around this very low temperature regime.
Therefore further analysis is necessary but we do not consider it in this article.

The analysis in the gapped phase is difficult since all the Wilson loop operators are excited and interacting each other through a constraint $\rho(\alpha)\ge 0$.
(The vertical shaded region in Figure \ref{fig phase diagram}.)
We can perturbatively analyze it just above the curve $\sqrt{-a/2b}=1/2$ by assuming that $|u_n|$ ($n\ge2$) are small \cite{Mandal:2009vz, Aharony:2003sx}.
Thus it is complicated to show the relation $\triangle \ge \mu$ in general.
On the other hand, if temperature or chemical potential is enough high, $|u_n|\sim 1$ is satisfied and we can evaluate the contribution of $A_0$ perturbatively.
There, analysis is possible as we will see in section \ref{Section high T} and we can guess the stability problem of the gapped phase through these results.

\subsection{$1/D$ corrections and problems in very low temperature regime}
\label{sec 1/D}
In this subsection, we evaluate the subleading $1/D$ corrections to the effective action (\ref{potential triangle}) in the low temperature and low chemical potential regime.
Then we will show that the $1/D$ expansion is not valid in a very low temperature regime.
After that we argue how the $1/D$ corrections modify the phase structure derived in the previous section.

We show the calculation of the $1/D$ corrections in appendix \ref{app 1/d} and, by using it, we obtain the relevant terms of the effective action as 
\begin{align} 
{\cal S}(\triangle, \{u_n\})/(DN^2)=C_0+ C_2  |u_1|^2 +C_4 |u_1|^4
+\cdots +O(1/D^2) ,
\label{s-eff-all}
\end{align} 
where
\begin{align} 
C_{0}=&-\frac{\beta \triangle^4}{8 \tilde{\lambda}
}+\frac{\beta \triangle}{2} \nonumber 
\\ &+\frac{\beta \triangle}{D}
\left[ \left( 1+\frac{\tilde{\lambda} }{4\triangle^3}
  \right)^{\frac{1}{2} } -1-\left(\frac{\tilde{\lambda}
  }{4\triangle^3} \right)-\frac{1}{4}\left(\frac{\tilde{\lambda}
  }{4\triangle^3} \right)^2 \right], 
\label{C0}
\end{align}
\begin{align}
C_{2}=&
\frac{1}{D}-x\left(k_1+k_2 \frac{y+y^{-1}}{2}  \right) +\frac{\beta \triangle}{D}x\left(k_1+k_2 \frac{y+y^{-1}}{2}  \right) \nonumber \\
& \times \Biggl[
  \left(\frac{\tilde{\lambda} }{4\triangle^3} \right) \left(
  1+\frac{\tilde{\lambda} }{4\triangle^3} \right)^{-\frac{1}{2} } +\frac{\frac{\tilde{\lambda}
    }{4\triangle^3}}{1+\frac{\tilde{\lambda} }{4\triangle^3} }
  -4\left(\frac{\tilde{\lambda} }{4\triangle^3}
  \right)-3\left(\frac{\tilde{\lambda} }{4\triangle^3} \right)^2
  \Biggr] +O(x^2),
\end{align}
\begin{align}
C_{4}=& \frac{\beta  \triangle}{2D} x^2 \left(k_1+k_2 \frac{y+y^{-1}}{2}  \right)^2 \left(\frac{\tilde{\lambda} }{4\triangle^3}
  \right)^2  \nonumber \\
 &\times \Biggl\{
\left[ -\frac12 \left( 1+\frac{\tilde{\lambda} }{4\triangle^3}
  \right)^{-\frac{3}{2} } -1 \right] 
+(2+\beta \triangle ) \left[
  -\frac{1}{
    \left( 1+\frac{\tilde{\lambda} }{4\triangle^3} \right)^2 }
  -2 \right] \Biggr\} \nonumber \\
&+\frac{\beta  \triangle}{2D} x^2 \left(k_2 \frac{y-y^{-1}}{2}  \right)^2 \left(\frac{\tilde{\lambda} }{4\triangle^3}
  \right)^2 \nonumber \\
&\times
\left\{
  \beta  \triangle \left[
  -\frac{1}{
    \left( 1+\frac{\tilde{\lambda} }{4\triangle^3} \right)^2 }
  -2 \right]  
-2 - \left(1+ \frac{\tilde{\lambda} }{4\triangle^3}\right)^{-3/2}\left(1+\frac{1}{2} \frac{\tilde{\lambda} }{4\triangle^3} \right)     \right\}+O(x^3) .
\end{align} 
Here $x \equiv e^{-\beta \triangle }$ and $y \equiv e^{-\beta \mu }$.
Note that higher order terms of $x$ are irrelevant in low temperature, since the transitions happen around $x \sim 1/D$ in $\mu=0$ case and the critical temperatures will decrease as $\mu$ increases.

Now we derive the effective action for the Wilson loop operators as in the previous section.
By solving the saddle point equation for $\triangle$,  
we obtain the condensation
\begin{align} 
\frac{ \triangle}{\tilde{\lambda}^{1/3} }=1+\frac{1}{D}
\left(\frac{7\sqrt{5}}{30}-\frac{9}{32}  
 \right)  +  \frac{2}{3}\bar{x}\left(k_1+k_2 \frac{y+y^{-1}}{2}  \right)  |u_1|^2 +\cdots ,
\label{condensation 1/D}
\end{align} 
where $\bar{x}\equiv e^{-\beta \tilde{\lambda}^{1/3} }$.
By substituting this solution into (\ref{s-eff-all}),
we obtain the effective action for the Wilson loops as
\begin{align}
{\cal S}/(DN^2)= \beta \tilde{\lambda}^{1/3} 
\epsilon_0 +  a' |u_1|^2 + b'|u_1|^4+\cdots,
\label{LG'}
\end{align}
with
\begin{align} 
\epsilon_0=&\frac{3}{8}+\frac{1}{D}
\left(-\frac{81}{64}+\frac{\sqrt{5}}{2}   \right), \label{free energy uni} \\  
a'=&\frac{1}{D}-\bar{x}\left(k_1+k_2 \frac{y+y^{-1}}{2}  \right) \left( 1+ \frac{\tilde{\lambda}^{1/3}\beta }{D} \left(
\frac{203}{160} -\frac{\sqrt{5}}{3} \right)\right)  ,
\label{a in 1/D}
\\ b'=& \bar{x}^2 \left(k_1+k_2 \frac{y+y^{-1}}{2}  \right)^2 \left[  \frac{\tilde{\lambda}^{1/3} \beta }{3}
+\frac{\tilde{\lambda}^{1/3} \beta }{D} \left(
\tilde{\lambda}^{1/3} \beta \left( \frac{229}{300}-\frac{2\sqrt{5}}{9} \right) +\frac{3181}{2400}-\frac{391\sqrt{5}}{1800}
\right) \right]   \nonumber \\
& - \frac{\tilde{\lambda}^{1/3} \beta}{D}\bar{x}^2 \left(k_2 \frac{y-y^{-1}}{2}  \right)^2
\left( 
\tilde{\lambda}^{1/3} \beta \frac{33}{400} +\frac{\sqrt{5}}{160} +\frac{1}{32}  
\right) . \label{b 1/D}
\end{align} 
From these expressions, we immediately find that the $1/D$ expansion in $T/\tilde{\lambda}^{1/3}<1/D$ regime is problematic, since several $1/D$ corrections involve  $\beta \tilde{\lambda}^{1/3}$ factors.
Thus if $\beta \tilde{\lambda}^{1/3} \sim D$ ($T/\tilde{\lambda}^{1/3} \sim 1/D$), these corrections become the same order to the leading terms.
Similar terms may arise in higher $1/D$ corrections also and the $1/D$ expansion will not be reliable in such a very low temperature regime.
Thus the $1/D$ expansion would be valid until $T/\tilde{\lambda}^{1/3} \sim D^{-\gamma}$,
where $\gamma$ is a positive constant\footnote{
$\gamma$ will be less than 1 through the arguments in this section.
In order to determine $\gamma$ precisely, we have to evaluate the higher order corrections of the $1/D$ expansion and it has not been done.}
\footnote{
In $\mu=0$ case, the very low temperature regime is in the uniform phase and physical quantities do not depend on $T$.
Thus this problem was not observed in \cite{Mandal:2009vz}.
However we cannot rule out a slight possibility that new phases appear in this regime.
Fortunately, numerical analyses show that such new phases do not arise \cite{Aharony:2005ew, Aharony:2004ig, Kawahara:2007fn, Azuma:2007fj, Azeyanagi:2009zf}. }.
This regime is depicted as the inclined shaded region in Figure \ref{fig phase diagram}.

Now we evaluate the effective action (\ref{LG'}) and argue the low temperature phase structure involving the $1/D$ corrections.
We can show that $b'$ is positive at $a'=0$.
Thus the argument in the previous section is still valid and the phase structure does not change.
The curve $a'=0$ and $\sqrt{-a'/2b'}=0$ give the two phase transition lines in the $\mu-T$ plane.
However we can see that the curve $a'=0$ ends at $\mu/\tilde{\lambda}^{1/3}=1$ on $T=0$ again.
This result is the same to the leading result in (\ref{eq mu-c1}).
This strange fact also indicates that the $1/D$ expansion is invalid in the very low temperature.

\subsection{General chemical potential}
\label{sec general mu}
Until now, we have studied the phase structure only for the simple chemical potential (\ref{simple mu}).
In this subsection, we take each $\mu_I$ arbitrary value and consider general chemical potentials.

Before evaluating the general chemical potentials, we discuss a small $k_2 \sim 1/D$ case, in which the contribution of the chemical potential will be comparable to the $1/D$ corrections derived in the previous subsection.
Even in this case, the phase structure in the low temperature and low chemical potential regime is similar.
On the curve $a'=0$, if $\mu$ sufficiently closes to $\tilde{\lambda}^{1/3} $, $k_2 e^{-\beta(\tilde{\lambda}^{1/3} -\mu)}$  will be dominant in (\ref{a in 1/D}) even if $k_2$ is small.
Thus the $\mu$ dependence is still large there.
The curve $\sqrt{-a'/2b'}=0$ also depends on $\mu$ strongly in a certain regime.
Therefore the qualitative nature of the phase structure is not modified.

Now we consider the general chemical potentials.
In this case, the one-loop contributions from the complex adjoint scalars are modified, and the second line of  the effective action (\ref{potential triangle}) becomes
\begin{align} 
&\frac{\beta \triangle}{2} -\sum_{n=1}^\infty 
\left[ e^{-n\beta \triangle}
\left(\frac{D-2 \lfloor D/2 \rfloor}{D} +\frac{2}{D}\sum_{I=1}^{\lfloor D/2 \rfloor}  \frac{e^{-n\beta \mu_I}+e^{n\beta \mu_I}}{2}\right)
\right] 
\frac{|u_n|^2}{n} .
\label{potential general mu}
\end{align} 
During the derivation of this potential, we have assumed $\triangle \ge \mu_I$ for all $\mu_I$.

Here we consider the phase structure.
As we have argued in the case of $k_2 \sim 1/D$, even if each contribution of the chemical potentials appears with the $1/D$ factor, the largest chemical potential will be dominant in some regimes to fix the phase structure.
Therefore we will obtain similar phase structure.
Besides we can confirm that the condensation satisfies $\triangle(\beta,\{\mu_I \})>\mu_I$ in the uniform and non-uniform phase. 

In principle, there is a possibility that a different phase structure appears, since we have deformed the potential through the chemical potentials.
However it does not happen.
Especially, in the leading order of the $1/D$ expansion, we can prove that the phase structure is determined by the $|u_n|$ independent terms of the effective action (\ref{potential triangle}). 
Since the $|u_n|$ independent terms of the potential (\ref{potential general mu}) are the same to the previous ones in (\ref{potential triangle}), the phase structure does not change.
We show it in appendix \ref{app LG}.\\

As we have mentioned in the introduction,  
it was supposed that the study of the low temperature thermodynamics of the gauge theories with the finite chemical potential is difficult because of the perturbative instability of the massless adjoint scalars.
However, in our study, the condensation $\triangle(\beta, \mu)$ of the adjoint scalars  protects our model from the instability and we can investigate the low temperature phase structure.
In the next section, we will explore the properties of the condensation in different regimes.

\section{Condensation in high temperature and high chemical potential regime}
\label{Section high T}
\subsection{Condensation in the large $D$ limit}
\label{sec high T}

We investigate the natures of our model in the high temperature regime and the high chemical potential regime.
In these regimes, the phase will be the gapped (deconfinement) phase and we can use an approximation $|u_n| \sim 1$ \cite{Yamada:2006rx, Mandal:2009vz, Aharony:2003sx}. 
Then we can use a perturbative analysis in $A_0$ around $A_0=0$.

In this subsection, we consider the simple chemical potential (\ref{simple mu}) and we
will show that a unique condensation $\triangle(\beta,\mu)>\mu$ exists for an arbitrary value of $\mu$ in the large $D$ limit.
This conclusion will be modified in the finite $D$ case as we will see in the next subsection.

In the large $D$ limit, we can ignore fluctuation of $A_0$ and it is enough to evaluate the saddle point equation (\ref{saddle point triangle}) only.
In $|u_n|=1$ case, this equation becomes
\begin{align} 
\frac{\triangle^3}{\tilde{\lambda} }=& 1 +2\sum_{n=1}^\infty 
e^{-n\beta \triangle}
\left[ k_1+k_2
\left(
 \frac{e^{-n\beta \mu}+e^{n\beta \mu}}{2} \right)
\right] \nonumber \\
=& 1 +\frac{2k_1}{e^{\beta \triangle}-1}+\frac{k_2}{e^{\beta (\triangle+\mu)}-1}
+\frac{k_2}{e^{\beta (\triangle-\mu)}-1}.
\label{saddle point high temperature}
\end{align}
The left hand side of this equation is simply increasing from 0 to infinity with respect to $\triangle$.
On the other hand, the right hand side is simply decreasing from infinity to 0 in $\triangle> \mu$ region.
As a result, this equation has an unique solution $\triangle(\beta,\mu)$ in $\triangle> \mu$ region.
(See Figure \ref{fig condensation}. Similar behaviour can be seen in $d=2$ and 3 dimensional gauge theories also.) 
There is another solution in $0<\triangle<\mu$. 
However this solution is unphysical since we have assumed $\triangle> \mu$ when we derived the saddle point equation (\ref{saddle point high temperature}).

Now we show solutions of (\ref{saddle point high temperature}) in several cases.
In $k_1=0$ case, if $\beta$ is sufficiently small, we obtain,
\begin{align} 
\triangle^2=&\frac{1}{2} \mu^2+\frac{1}{2} \sqrt{\mu^4+8\frac{\tilde{\lambda}}{\beta}}
& ( \beta \tilde{\lambda}^{1/3}\ll 1 )  .
\end{align} 
In $k_1 \ne 0$ case, if $\beta$ and $\mu$ are small, we obtain
\begin{align} 
\triangle=&\left( \frac{2\tilde{\lambda}}{\beta}\right)^{1/4} 
\left(1  +\frac{k_2 \mu^2}{4} \sqrt{\frac{\beta}{2\tilde{\lambda} } } \right)+\cdots  & (\mu^4 \beta /\tilde{\lambda}  \ll 1,~ \beta \tilde{\lambda}^{1/3}\ll 1 ).
\label{cond-high-T}
\end{align} 
These two solutions are valid in very high temperature.
If $\triangle$ is close to $\mu$, we obtain
\begin{align} 
\triangle=& \mu + \frac{ k_2 \tilde{\lambda} }{\beta \mu^3} +\cdots  & (\mu/\tilde{\lambda}^{1/3} \gg (k_2 T/\tilde{\lambda}^{1/3})^{1/4} ).
\label{cond-high-mu}
\end{align} 
This solution is valid in very high chemical potential regime.
Note that this condensation induces a light effective mass $\triangle^2-\mu^2 \sim 2 k_2 \tilde{\lambda} /\beta \mu^2$ for $\Phi^I$ in (\ref{phi kinetic}).
In the next subsection, we will see that such light mass modes cause a divergence in $1/D$ corrections and the $1/D$ expansion does not work there.
Thus the arguments in the very high chemical potential regime will be valid only in the $D\to \infty$ case.

\subsection{$1/D$ correction in very high chemical potential regime}
\label{sec high mu 1/D }
In this subsection, we evaluate the $1/D$ corrections in the very high chemical potential regime ($\mu/\tilde{\lambda}^{1/3} \gg (k_2 T/\tilde{\lambda}^{1/3})^{1/4} $), in which the equation (\ref{cond-high-mu}) is satisfied, and argue the validity of the $1/D$ expansion.

In the very high chemical potential regime, since $\beta(\triangle-\mu)$ will be small and $u_n$ will be close to $1$, the zero-mode of the Matsubara frequencies of $\Phi^I$ will be dominant.
Therefore the relevant $1/D$ corrections arise from loops of $\Phi^I$ and $b_{ab}$ and
the path integral of $A_0$.

First we evaluate the $1/D$ correction from $A_0$.
In the very high chemical potential regime, the dominant $\triangle$ dependent terms in (\ref{log det}) can be evaluated as
\begin{align} 
\frac{1}{2}\log\left\{
\left(\alpha_j-\alpha_i\right)^4+2\left(\triangle^2+\mu^2\right)\left(\alpha_j-\alpha_i\right)^2+\left(\triangle^2-\mu^2\right)^2\right\} .
\end{align} 
Then the effective action for $A_0$ is given by
\begin{align} 
\sum_{i,j}\frac{\tilde{D}}{2}\log\left\{
\left(\alpha_j-\alpha_i\right)^4+2\left(\triangle^2+\mu^2\right)\left(\alpha_j-\alpha_i\right)^2+\left(\triangle^2-\mu^2\right)^2\right\} 
-\frac{1}{2} \log(\alpha_j-\alpha_i)^2,
\label{log det high mu}
\end{align} 
where the last term is derived from (\ref{FP}) by assuming that $\alpha_i$ are small.
Now we assume $\triangle^2-\mu^2 \gg (\alpha_j-\alpha_i)^2$.
Then we can expand the first $\log$ term and obtain\footnote{If we start an assumption $\triangle^2-\mu^2 \ll (\alpha_j-\alpha_i)^2$ in (\ref{log det high mu}), the attractive force between $\alpha_i$ will be quite strong and this assumption will not be satisfied.} 
\begin{align} 
N^2\tilde{D}\log
(\triangle^2-\mu^2) 
+
2N\tilde{D}
\frac{(\triangle^2+\mu^2)}{(\triangle^2-\mu^2)^2
}\sum_{i=1}^N\alpha_i^2 
-\sum_{i,j}\frac{1}{2} \log(\alpha_j-\alpha_i)^2. 
\end{align} 
Note that this action is just a gaussian in $A_0$,
\begin{align} 
N^2\tilde{D}\log
(\triangle^2-\mu^2) 
+
2\tilde{\lambda}k_2
\frac{(\triangle^2+\mu^2)}{(\triangle^2-\mu^2)^2
}\Tr A_0^2/g^2. 
\end{align} 
Since the coefficient of $A_0^2$ will be enough large in $\mu/\tilde{\lambda}^{1/3} \gg (k_2 T/\tilde{\lambda}^{1/3})^{1/4} $ regime, $\alpha_i$ will be strongly trapped around $\alpha_i=0$.
Thus the assumption $\triangle^2-\mu^2 \gg (\alpha_j-\alpha_i)^2$ will be satisfied.
Then the gaussian integral of $A_0$ gives a $1/D$ correction and the effective action for $\triangle$ in the very high chemical potential regime will become
\begin{align} 
S_{eff}(\triangle)/DN^2=&
-\frac{\triangle^4}{8\tilde{\lambda}T} 
+\frac{ \tilde{D}}{D}\log
(\triangle^2-\mu^2) 
-\frac{1}{2D}\log\left( 
\frac{(\triangle^2-\mu^2)^2}{(\triangle^2+\mu^2)}\right) \nonumber \\
&+\left(O\left( \frac{1}{D} \right)  ~\text{from matter loops} \right) + \cdots .
\label{1/d A high mu}
\end{align} 
Here the first and the second terms are from (\ref{potential triangle}) with an approximation $\beta(\triangle-\mu) \ll 1$.
The third term is from the $A_0$ integral.
Thus the $1/D$ correction from $A_0$ is qualitatively not important for large $\tilde{D}$ case.
However, if $\tilde{D}=1$, this correction cancels the second term.
As a result, the arguments in the previous section will not be valid and (\ref{cond-high-mu}) will not be satisfied.
Thus the assumption $\triangle \sim \mu$ in $\mu/\tilde{\lambda}^{1/3} \gg (k_2 T/\tilde{\lambda}^{1/3})^{1/4} $ is not ensured.  
Therefore a different analysis is necessary in $\tilde{D}=1$ case and we will discuss it in the next section.

Now we evaluate the subleading $1/D$ corrections from the matter loops.
In the large $\tilde{D}$ case, we have confirmed that $A_0$ is sufficiently small and we can ignore it in the loop calculation at this order.
Then the dominant contributions of the matter loops are from the zero modes of the Matsubara frequencies of $\Phi^I$ and $b_{ab}$.
Therefore we can evaluate them by using a zero dimensional reduced model
\begin{align} 
S_{0d}=
-\frac{1}{4}\tilde{b}_{ab}M_{ab,cd}^{-1}\tilde{b}_{cd}+
\sum_{I=1}^{\tilde{D}}
\left( 
(\triangle^2-\mu^2) \tilde{\Phi}_a^{\dagger I} \tilde{\Phi}_a^I 
+g\sqrt{T} \tilde{b}_{ab}\tilde{\Phi}_a^{\dagger I} \tilde{\Phi}_b^I
\right) .
\label{zero dim matrix model}
\end{align} 
Here $\tilde{b}_{ab}$ and $\tilde{\Phi}^I_a$ are the zero-modes of the one dimensional fields.
Then we can calculate the $1/D$ corrections of the effective action (\ref{1/d A high mu}) as in appendix B of \cite{Mandal:2009vz},
\begin{align} 
\frac{1}{D}\left(
-\frac{k_2\tilde{\lambda}T}{(\triangle^2-\mu^2)^2} 
-\frac{1}{2} \left(\frac{ k_2 \tilde{\lambda}T}{(\triangle^2-\mu^2)^2} \right)^2 
-\frac{1}{2}
\sum_{m=1}^\infty
\frac{1}{m}  \left( -\frac{k_2\tilde{\lambda}T}{(\triangle^2-\mu^2)^2}  \right)^m
\right)  .
\label{1/d matter high mu}
\end{align} 
From this expression, we notice that if $(\triangle^2-\mu^2)^2 < k_2 \tilde{\lambda}T$, the last sum does not converge.
It means that the $1/D$ expansion does not work in this regime.
From (\ref{cond-high-mu}), it will happen 
\begin{align} 
\mu^4 > 4 k_2 \tilde{\lambda}T.
\label{bound high mu}
\end{align} 
Note that this estimate is crude, since, if $\mu$ is not sufficiently larger than $( k_2 \tilde{\lambda}T)^{1/4}$, the reduced model analysis (\ref{zero dim matrix model}) is not valid.
Especially in the uniform and the non-uniform phases, the contribution of the gauge field is relevant and the $1/D$ expansion still works as we have discussed in section \ref{sec 1/D} even if (\ref{bound high mu}) is satisfied.
By considering it, we conclude that the $1/D$ expansion is not valid in the horizontal shaded region in Figure \ref{fig phase diagram} schematically. 
Although the $1/D$ expansion does not work in this regime, 
we cannot conclude that the system is unstable there.
The divergence of the $1/D$ correction in (\ref{1/d matter high mu}) arises from the loops of the light mass modes of $\Phi^I$ and it is not clear whether it indicates an instability of the system or not.

\paragraph{$1/D$ correction in finite chemical potential}
Here we consider the $1/D$ corrections in the finite chemical potential regime $\mu/\tilde{\lambda}^{1/3} < (k_2 T/\tilde{\lambda}^{1/3})^{1/4} $. 
In a finite temperature, the calculation for the corrections will be complicated but, if temperature is enough high, a zero dimensional analysis similar to (\ref{zero dim matrix model}) is possible.
Then it is not difficult to show that the $1/D$ corrections converge if $(\triangle^2-\mu^2)^2 > k_2 \tilde{\lambda}_0 T$ is satisfied.
Therefore we can guess that the $1/D$ expansion is valid in the finite temperature case also.

By using the $1/D$ expansion method, we have revealed that, if the chemical potential is not very high, our model in the high temperature regime and the low temperature regime is stable.
The stability in both of the regimes will support the stability of the unknown regime in the gapped phase. 
(The vertically shaded region in Figure \ref{fig phase diagram}.)
However we have investigated only one condensate vacuum and there is a possibility that this vacuum is just a local minimum and more stable vacua exist\footnote{The appearance of other phases might be natural. 
If $\mu$ is large, $ \langle \Tr | \Phi^I |^2 \rangle \gg \langle \Tr Y^{i2} \rangle$ will be satisfied since the chemical potential makes the effective masses of $\Phi^I$ light. Then the eigenvalue distribution of the adjoint scalars will be pancake like, and, intuitionally, a doughnut like distribution may be favoured.  
If such a transition happens, it may correspond to the Meyers-Perry black hole/black ring transition in higher dimensional gravity \cite{Emparan:2008eg}.
Besides, several intermediate deformed Meyers-Perry black hole solutions also exist in gravity \cite{Dias:2009iu}.
However, we have not found such new phases in our model and it is interesting to investigate them further.
}.
Another possibility is that the model is unbounded below in the finite chemical potential.
Thus we conclude that our model is at least meta-stable if the chemical potential is not very high.

\subsection{Condensation in $\tilde{D}=1$ case}
\label{sec D=1}
As we have seen in equation (\ref{1/d A high mu}), the $\tilde{D}=1$ case is special.
In this subsection, we show that a consistent condensation will happen in $\tilde{D}=1$ case also by using a different analysis up to $1/D$ order.
However, this condensation will give rise to a small effective mass $\triangle^2-\mu^2 $ in a very high chemical potential and it will cause a divergence in the next order of the expansion.
This is similar to the behaviour of the $\tilde{D}>1$ case and the $1/D$ expansion will not work  in the very high chemical potential regime.
In addition, we will show that our model has complex fuzzy sphere like saddle points in $\tilde{D}=1$ case, although their physical interpretations have not been understood.

For simplicity, we consider a sufficient high temperature regime ($\beta \tilde{\lambda}^{1/3} \ll 1$) only.
Then the zero-modes of the Matsubara frequencies will be dominant and the model reduces to a zero dimensional model,
\begin{align} 
S = 
&
 \Tr 
\left(
-2\mu i g_0 
\Phi^{\dagger } \left[ A_0,  \Phi\right] -\mu^2 \Phi^{\dagger }\Phi
\right)
 \nonumber \\
&+ \sum_{i,j=1}^{D-2}g^2_{0} \left(\Phi^{\dagger }_a \Phi_b +\frac{1}{2} Y^{i}_a Y^{i}_b +\frac{1}{2}A_{0a}A_{0b}\right)  
M_{ab,cd}
\left(\Phi^{\dagger }_c \Phi_d +\frac{1}{2} Y^{j}_c Y^{j}_d +\frac{1}{2}A_{0c}A_{0d}\right),
\label{action 0d}
\end{align} 
where we have scaled matrices appropriately and the coupling is defined as $g_0^2=g^2T$.
By employing an auxiliary matrix $B_{ab}=\triangle^2 \delta_{ab}+g_0b_{ab}$ where $b_{ab}$ satisfies $b_{aa}=0$, this action becomes,
\begin{align} 
S = 
&
 \Tr 
\left(
-2\mu  g_0 
W \left[ A_0,  X \right]+\frac{\triangle^2 -\mu^2}{2} 
\left(X^2+W^2 \right) 
+\frac{\triangle^2}{2}A_0^2 + \sum_{i=1}^{D-2} \frac{\triangle^2}{2}Y^{i2} 
\right) \nonumber \\
&
-\frac{DN^2 \triangle^4}{8\tilde{\lambda}_0} -\frac{1}{4}b_{ab}M_{ab,cd}^{-1}b_{cd} +\frac{g_0}{2}b_{ab}\left(Y^{i}_a Y^{i}_b+X_a X_b+W_aW_b  +A_{0a}A_{0b}\right).
\end{align} 
Here we have used $\Phi=(X+iW)/\sqrt{2}$ and $\tilde{\lambda}_0=g_0^2 ND$.
We evaluate this action up to the second order of the $1/D$ expansion.
In this case, we can ignore the interactions between $A_0, X,W$ and $b_{ab}$ in the last term. 
By integrating out $Y^i$ and $b_{ab}$ through a technique in \cite{Mandal:2009vz}, we obtain
\begin{align} 
S = 
& \Tr 
\left(
-2\mu  g_0 
W \left[ A_0,  X \right]+\frac{\triangle^2 -\mu^2}{2} 
\left(X^2+W^2 \right) 
+\frac{\triangle^2}{2}A_0^2 \right) \nonumber \\
&+DN^2\Biggl[
-\frac{ \triangle^4}{8\tilde{\lambda}_0} 
+\frac{k_1}{4} \log \triangle^4 \nonumber \\
&+\frac{1}{D} \left(-\frac{k_1 \tilde{\lambda}_0}{\triangle^4} -\frac{1}{2}\left(\frac{k_1 \tilde{\lambda}_0}{\triangle^4} \right)^2
+\frac{1}{2} \log\left(1+\frac{k_1 \tilde{\lambda}_0}{\triangle^4} \right)     \right)    
+O\left(\frac{1}{D^2}  \right) 
\Biggr].
\label{action 0d 2}
\end{align}
Here $k_1=(D-2)/D$.

Now we evaluate the path integral of $X,W$ and $A_0$ and derive an effective action for $\triangle$.
A formula for the following three matrix model is available \cite{Kazakov:1998ji}:
\begin{align}
&S = \Tr \left\{aM_1[M_2,M_3] + \frac{b}{2}\left(M_1^2+M_2^2 \right)   +\frac{c}{2} M_3^2 \right\}, \nonumber \\ 
Z=&\int DM_1 DM_2 DM_3 \exp\left(-S  \right) \nonumber \\
  =& C a^{-N^2}  \int d m_1 \cdots d m_N \prod_{i \ne j}\frac{m_i - m_j}{ m_i - m_j+1 } \prod_{i} e^{-\lambda_M m_i^2} \nonumber \\
=  & C a^{-N^2} e^{-N^2 F_0(\lambda_M)}+\cdots,
\end{align} 
where $a,b,c$ are constants and $\lambda_M \equiv N/g_M^2 = cb^2/2a^2$.
$C$ is an irrelevant factor.
 We have ignored $1/N$ corrections.
Here the free energy is given by
\begin{align} 
F_0(\lambda_M) & \to -\frac{1}{2}\log g_M^2 +\frac{1}{2}g_M^2 + \cdots   &(g_M \rightarrow 0)  ,
\label{small g_m} \\
&\to \frac{3(12\pi)^{2/3}}{40}g_M^{-2/3}+O(g_M^{-5/3})  &(g_M \rightarrow \infty) .
\end{align} 
In our case, from (\ref{action 0d 2}), $g_M$ becomes
\begin{align} 
g_M^2=\frac{8\mu^2 \tilde{\lambda}_0}{\triangle^2(\triangle^2-\mu^2)^2D} .
\end{align} 
If $\mu$ is small, $g_M$ will be also small.
Then (\ref{small g_m}) gives us standard $\log$ terms plus $O(1/D^2)$ corrections.
Thus, in the small $\mu$ case, a consistent condensation $\triangle>\mu$ will happen as usual.
On the other hand, if $\mu$ is large and $\triangle^2-\mu^2 $  is small, then $g_M^2$ will be large.
In this case, we obtain the effective action for $\triangle$ as
\begin{align} 
\frac{S(\triangle)}{DN^2}=&-\frac{\triangle^4}{8\tilde{\lambda}_0}
+\frac{k_1}{4} \log \triangle^4
+ \frac{3(12\pi)^{2/3}}{40D}
\left(
  \frac{\triangle^2(\triangle^2-\mu^2)^2 D}{8\mu^2 \tilde{\lambda}_0}  
\right)^{1/3} 
\nonumber \\
&
+\frac{1}{D}\left(-\frac{k_1 \tilde{\lambda}_0}{\triangle^4} -\frac{1}{2}\left(\frac{k_1 \tilde{\lambda}_0}{\triangle^4} \right)^2
+\frac{1}{2} \log\left(1+\frac{k_1 \tilde{\lambda}_0}{\triangle^4} \right)     \right)    
+\cdots.
\label{D=1 effective action}
\end{align}
From this action, we can derive a saddle point equation for $\triangle$ and we can obtain a unique saddle point $\triangle(\mu)$ which satisfies $\triangle(\mu)-\mu \gtrsim 0$ for any $\mu$.

Therefore, up to this order of the $1/D$ expansion, we obtain the consistent condensation and the system is stable even in $\tilde{D}=1$ case.
We consider the high temperature approximation in this section but similar analysis will be possible in a finite temperature also.
However, since $\triangle^2-\mu^2$ will be small in the very high chemical potential regime, the next order terms will diverge and the $1/D$ expansion will not be valid as in the large $\tilde{D}$ case.
(The appearance of the fractional power of $D$ in the third term of (\ref{D=1 effective action}) also implies a problem of the $1/D$ expansion in this regime.)

\paragraph{Fuzzy solutions?}
Now we discuss possible saddle points of the zero dimensional action (\ref{action 0d 2}).
Since the cubic interaction in (\ref{action 0d 2}) can be regarded as a CS like term, the action has a complex fuzzy sphere like saddle point \cite{Ishiki:2010pe},
\begin{align} 
X = -i \frac{\sqrt{\triangle^2(\triangle^2-\mu^2)}}{2\mu g} J_1,  ~W =-i \frac{\sqrt{\triangle^2(\triangle^2-\mu^2)}}{2\mu g} J_2,  
~A_0= -i \frac{\triangle^2-\mu^2}{2\mu g} J_3, 
\end{align}
where $J_i$ are the generators of the $N$ dimensional irreducible representation of $SU(2)$, which satisfy $[J_i,J_j]=i \epsilon_{ijk}J_k$.
By replacing $J_i$ with reducible representations, we can obtain many saddle points.
However, these fuzzy sphere like solutions are not hermitian\footnote{
At these saddle points, $\triangle<\mu$ may not be forbidden.
(However it will be unstable.)
Then $X$ and $W$ can be hermitian but $A_0$ is still not.
} and
 physical interpretation is unclear.

In addition to these complex saddle points, it might be possible to find different fuzzy solutions in (\ref{action 0d}) or (\ref{action 0d 2}) as in the studies in \cite{Iso:2001mg, Ishiki:2010pe, Jatkar:2001uh, Kimura:2001uk, Azuma:2004zq}.

\paragraph{General chemical potentials}
Now we consider the general chemical potentials as in section \ref{sec general mu}.
If the values of several chemical potentials are the same and larger than the others, the analysis in the previous section is valid.
If only one chemical potential is very large, we can approximately apply the analysis in this section by ignoring other chemical potentials.
Thus the consistent condensation $\triangle>\mu_I$ will always happen up to the $1/D$ order.

\section{High temperature condensation in higher dimensional gauge theory}
\label{sec high d}
In this section, we consider the generalization of our argument about the condensation to $d$ dimensional gauge theory,
\begin{align} 
S = \int_0^{\beta} dt \int d^{d-1}x \,
\Biggl[
&
 \Tr 
\left(
\frac{1}{4g^2_{d}} F_{\mu\nu}^2
-
\sum_{I=1}^{\tilde{D}}\Phi^{\dagger I} \left( (D_0-\mu)^2 +D_i^2 \right)  \Phi^{I}
- \sum_{i=2\tilde{D}+1}^D
\frac12  Y^{i} D_{\mu}^2  Y^{i}
\right)
  \nonumber \\
& + \sum_{I,J,i,j}g^2_{d} \left(\Phi^{\dagger I}_a \Phi^{I}_b +\frac{1}{2} Y^{i}_a Y^{i}_b \right)  
M_{ab,cd}
\left(\Phi^{\dagger J}_c \Phi^{J}_d +\frac{1}{2} Y^{j}_c Y^{j}_d \right) \Biggr].
\label{action high d}
\end{align} 
Here $g_{d}$ is the gauge coupling.
We consider the simple chemical potential (\ref{simple mu}).
We will show that the condensation of the adjoint scalars can happen at least in the high temperature regime in the leading order of the $1/D$ expansion. 

First we employ an auxiliary field $B_{ab}$ as in the $d=1$ case.
Then we assume that this field condenses as $B_{ab}=\triangle^2\delta_{ab}$, where $\triangle$ does not depend on the time and position.
We also assume that this condensation satisfies $\triangle > \mu$.
Under these assumptions, we can exactly derive a saddle point equation to determine the condensation $\triangle$ in the large $D$ limit.

In the large $D$ limit, we can ignore the contribution of the spatial components of the gauge field $A_i$ by regarding $D \gg d-1$.
The interactions between $b_{ab}$ and $\Phi^I$ and $Y^i$ are also suppressed. 
(Here $b_{ab}$ is defined as in (\ref{fluct}) and satisfies $\int dt d^{d-1}xb_{aa}=0 $.)
In the high temperature regime, we can also ignore $A_0$. 
Then, we can integrate out $\Phi^I$ and $Y^i$ and obtain the effective action for $\triangle^2$ as in section \ref{sec Eff}.
From this effective action, the saddle point equation is obtained as
\begin{align} 
  \frac{\triangle^2}{\tilde{\lambda}_{d} T} =&
  \sum_{n} \int \frac{ d^{d-1}p}{(2\pi)^{d-1}} \Biggl[ \frac{2k_1}{\left(\frac{2\pi n}{\beta}   \right) ^2 + \vec{p}^2+\triangle^2 }  \nonumber \\
&+\frac{k_2}{\left(\frac{2\pi n}{\beta}  +i \mu  \right) ^2 + \vec{p}^2+\triangle^2 } 
+ \frac{k_2}{\left(\frac{2\pi n}{\beta}  -i \mu  \right) ^2 + \vec{p}^2+\triangle^2 } \Biggr].
\end{align} 
Here $\tilde{\lambda}_{d} \equiv g_d^2 N D$  is the $d$ dimensional 'tHooft like coupling.
Generally solving this equation is still complicated
and we evaluate it by taking a high temperature limit.
Here the zero modes of the Matusbara frequencies are dominant and we can ignore the non-zero modes.
Then the equation is simplified as,
\begin{align} 
\frac{(2\pi)^{d-1}}{2V_{d-2}}
\left( \frac{\Lambda^{5-d}}{ \tilde{\lambda}_{d} T} \right) 
\left(  \frac{\triangle^2}{\Lambda^2}\right) 
=&
k_1 f_{d} (\triangle^2/\Lambda^2)+ k_2 f_{d} ((\triangle^2-\mu^2)/\Lambda^2),
\label{condensation high d}
\end{align} 
where
\begin{align} 
f_{d}(x) \equiv \Lambda^{3-d}  \int_0^\Lambda dp \frac{ p^{d-2}
}{ p^2+x \Lambda^2 }.
\end{align} 
Here $V_{d-2}$ is the volume of the $d-2$ dimensional unit sphere and $\Lambda$ is a momentum cut off.
$T$  appears only through $\tilde{\lambda}_{d}T$ in this equation.
Thus we can regard $\tilde{\lambda}_{d}T$ as an effective coupling.
This equation determines $\triangle$ in terms of the effective coupling $\tilde{\lambda}_{d}T$, $\mu$ and the cut off $\Lambda$.  
We tune $\Lambda$ dependence of $\tilde{\lambda}_{d}T$ such that the condensation $\triangle$ is fixed for a particular physical value at a certain chemical potential $\mu$.

\FIGURE{
\includegraphics[scale=0.75]{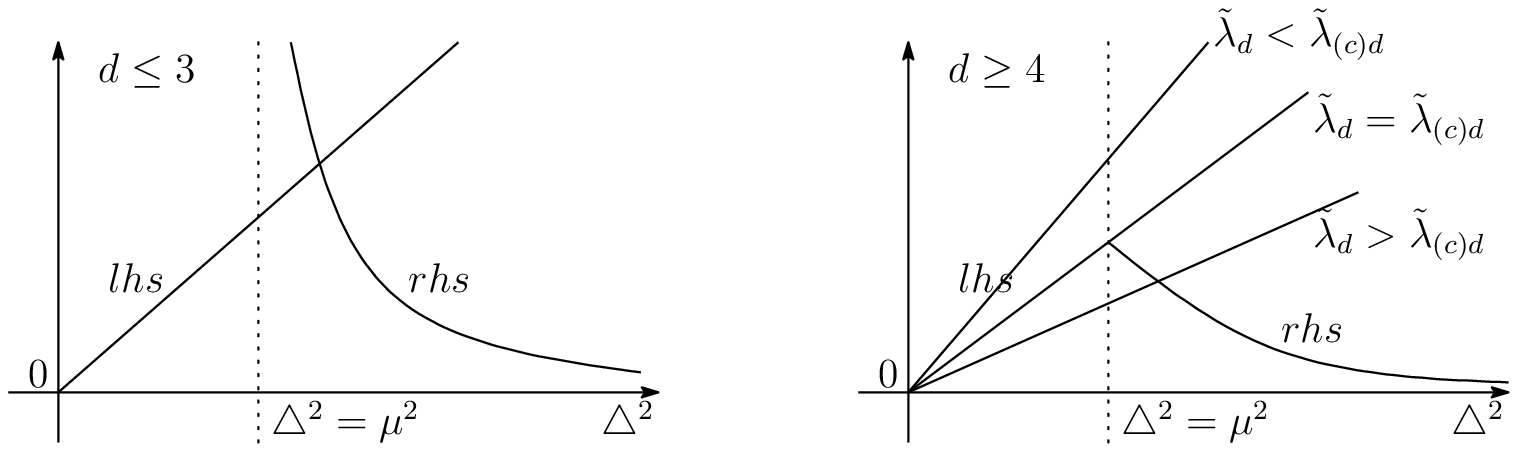}
\caption{Solution of (\ref{condensation high d}).
In $d\le 3$ case (the left plot), the curve represents the schematic behaviour of the right hand side of (\ref{condensation high d}) in $\triangle^2>\mu^2$ region.
It diverges at $\triangle^2=\mu^2$.
The straight line is the left hand side of (\ref{condensation high d}) and the crossing point gives the solution of (\ref{condensation high d}).
Note that a unique solution exists for any value of $\mu$.
In $d\ge 4$ case (the right plot), the curve does not diverge at $\triangle^2=\mu^2$.
As a result the curve cannot cross the line for small $\tilde{\lambda}_{d} $ (or large $\mu$) and  the solution does not exist in this case.
 }
\label{fig condensation}}

Now we evaluate the equation (\ref{condensation high d}).
We show the explicit expressions for $f_d(x)$ in appendix \ref{app fd} and we can derive the condensation by solving it.
Instead of doing it, here we argue the condition for the existence of the condensation in $\triangle> \mu$ region through the qualitative properties of $f_d(x)$.
We can show that $f_d(x)$ is simply decreasing in $x>0$ and behaves as
\begin{align} 
f_d(x)
\left\{
 \begin{array}{lll}
\to \infty & x \to +0  &(d \le 3)\\
\to 1/(d-3) & x \to +0 &(d \ge 4)\\
\to 0 & x \to +\infty & 
\end{array}
\right. .
\label{fd}
\end{align} 
Thus the right hand side of (\ref{condensation high d}) diverges at $\triangle=\mu$ in the $d\le 3$ case but does not in the $d\ge 4$ case.
Here the consistent solution of (\ref{condensation high d}) is given by a crossing point in Figure \ref{fig condensation}.
Therefore the equation (\ref{condensation high d}) always has a unique consistent solution in $d\le 3$.
On the other hand, in $d \ge 4$ case,  
one solution exists only if the following equation is satisfied,
\begin{align} 
\frac{ \tilde{\lambda}_{d} T}{\Lambda^{5-d}}
\ge
\frac{(2\pi)^{d-1}}{2V_{d-2}}
 \left( \frac{\mu}{\Lambda}\right)^2 
\frac{1}{k_1 f_{d} (\mu^2/\Lambda^2)+ k_2/(d-3) }.
\label{critical coupling}
\end{align} 
We have obtained this condition by evaluating (\ref{condensation high d}) at $\triangle=\mu$.
Thus the consistent condensation does not happen if the effective coupling is weak.
Equivalently we can say that, for a given coupling, a critical chemical potential $\mu_c$ exists, which saturates (\ref{critical coupling}), and the condensate happens only if $\mu<\mu_c$.
We summarize this result in Figure \ref{fig large d phase}.

\FIGURE{
\includegraphics[scale=0.75]{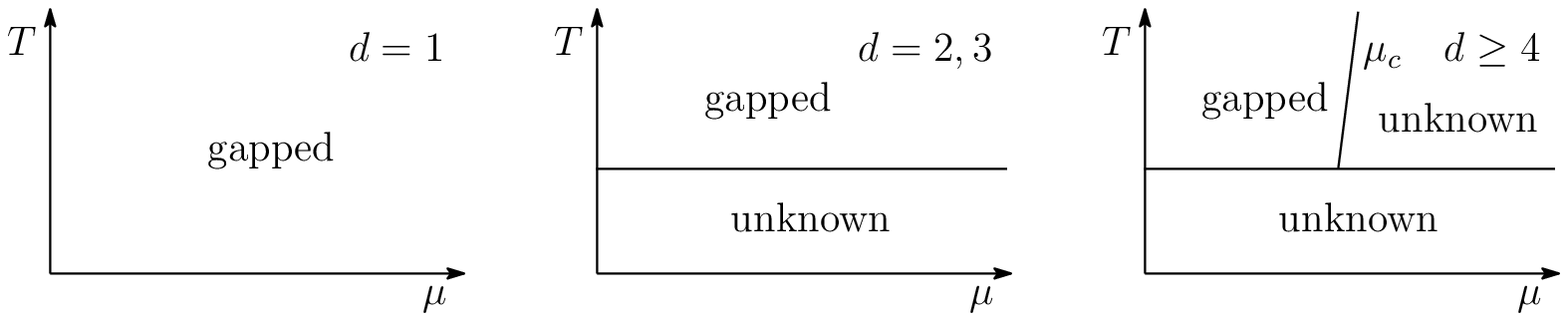}
\caption{Schematic phase diagrams of the $d$ dimensional large $N$ gauge theories in $\mu-T$ space at $D=\infty$.
In the $D=\infty$ case, the problems from the $1/D$ corrections do not arise.
In $d=1$ case, since the two critical temperatures become $0$ at $D=\infty$, only the gapped phase appears. (See (\ref{Tc1}) and (\ref{Tc2}).)
In $d \ge 2$ case, we evaluated only the high temperature regime.
In  $d \ge 4$ case, there is a critical chemical potential and the condensation does not happen beyond it.
The nature of this regime has not been understood.
}
\label{fig large d phase}
}

One important unsolved issue is the nature of $\mu>\mu_c$ region, in which the condensation does not happen.
One possibility is that the system is destabilized by the chemical potential.
Another possibility is that the system is still stable but described by a highly interacting theory.

Although the above results are exact in the large $D$ limit,
as we have observed in section \ref{sec high mu 1/D }, the $1/D$ corrections are important in the very high chemical potential regime in finite $D$ case.
Thus our results may be valid only in the $D \to \infty$ case in this regime.

Even if we start the supersymmetric gauge theory\footnote{How to take large $D$ limit in the supersymmetric gauge theory is difficult problem, since the number of the bosons and fermions grow at different rates as $D$ grows large.
By taking the high temperature limit and ignoring fermions, we can ignore this problem.}, the behaviour will be the same in the high temperature regime, since all the fermions are decoupled if the temperature is sufficiently high.
Therefore the analysis in this section may be valid in the supersymmetric gauge theory also.

\section{Conclusions and discussions}
\label{Conclusion}
In this article, we investigated the thermodynamics of the large $N$ gauge theories with the chemical potentials.
Because of the condensation of the adjoint scalars in the large $D$, we can analyze the effect of the finite chemical potential even in the perturbatively massless gauge theories.
It is an important step towards understanding the phase structure of large $N$ gauge theories.
However the light mass modes of the adjoint scalars in the very high chemical potential regime cause the $1/D$ expansion to diverge.
Understanding of the nature of this divergence is important to figure out whether the system is stable or not in this regime for finite $D$.
In addition, if we can understand the $1/D$ corrections in the higher dimensional gauge theories, it may be possible to apply our analysis to D brane theories at a high temperature.
It would then be interesting to compare our results in the weak coupling regime to the strong coupling results predicted by the dual gravity analysis \cite{Harmark:1999xt}.
\\

We also studied the phase structure of the one-dimensional gauge theory as in Figure \ref{fig phase diagram}.
As we have mentioned in the introduction, this theory is related to D1 branes on a small circle and it is interesting to evaluate this system by using the dual gravity and compare each other.

However we have not completed the understanding of the phase structure of our model.
There is a possibility that other phases appear.
If such phases are found, they may correspond to non-spherical black objects like a black ring or hairy black holes in the dual gravity.

In the sufficient high temperature regime, the one-dimensional model reduces to the zero-dimensional model (\ref{action 0d}). 
This model is similar to the bosonic IKKT matrix model with negative mass and imaginary CS like terms.
As we have discussed in section \ref{sec D=1}, fuzzy solutions might exist in this model \cite{Iso:2001mg, Ishiki:2010pe, Jatkar:2001uh, Kimura:2001uk, Azuma:2004zq}.
Therefore, exploring this matrix model may be the simplest way to find the new phases.\\

Our phase structure in the bosonic gauge theory is quite different from the results in the one-dimensional supersymmetric gauge theory predicted from D0 brane black hole \cite{Itzhaki:1998dd, Harmark:1999xt}.
There, the low temperature phase transition does not happen and the system is always in the gapped (deconfinement) phase.
Besides if the chemical potential is larger than a critical chemical potential $\mu_c=c T$, the system is destabilized.
These differences indicate that the contribution of the fermions is relevant in the low temperature regime.
One possibility is that, in the supersymmetric theory, the effective action (\ref{potential triangle}) is modified such that the condensation behaves as $\triangle(T,\mu) \sim T$ in the low temperature regime.
Then the potentials for the Wilson loop operators may be unstable at $|u_n|=0$ even around $T=0$ and the gapped (deconfinement) phase will be preferred.
In addition, the system will be destabilized if $\mu > \triangle \sim T$.
However, how to take a large $D$ limit in supersymmetric theories has not been understood and we cannot show such a mechanism yet.

\paragraph{Acknowledgements }
I would especially like to thank Gautam Mandal for useful discussions and for several detailed comments on the manuscript.
I would also like to thank Avinash Dhar, Oleg Evnin, Shoichi Kawamoto, Hiroaki Kohyama, Manavendra Mahato, Shiraz Minwalla, Shinji Shimasaki, Sandip Trivedi and Spenta Wadia for useful discussions.

\appendix

\section{Comment on phase structure in $1/D$ expansion}
\label{app LG}

As we have seen in section \ref{Section low T}, the effective action for $u_n$ determines the phase structure of the one-dimensional large $N$ gauge theory.
Especially, the potential for $|u_1|$ is relevant in the low temperature and low chemical potential regime \cite{Aharony:2003sx, AlvarezGaume:2005fv}.
In this appendix, we argue how the phase structure is fixed in the $1/D$ expansion.

Generally, the relevant terms in the effective potential for the condensation $\triangle^2 \equiv \tau$ and $u_1$ will be given as
\begin{align} 
S(\tau, u_1)/DN^2=C_0(\tau)+C_2(\tau) |u_1|^2+C_4(\tau) |u_1|^4+\cdots.
\label{eff-action-gen}
\end{align} 
Thus the saddle point equation for $\tau$ is
\begin{align} 
C'_0(\tau)+  C'_2(\tau) |u_1|^2+  C'_4(\tau) |u_1|^4=0.
\end{align} 
Here "$'$" denotes $\partial/\partial \tau $.
We assume that $C'_2$ and $C'_4$  are small compare to $C'_0$.
Indeed, by estimating the $1/D$ corrections, we will see that $C_0$, $C_2$ and $C_4$ are order $1$, $1/D$ and $1/D^3$ respectively near the critical point. 
Then we can solve this equation perturbatively by expanding $\tau=\tau_0+\tau_1+\cdots$.
First $\tau_0$ is given by a solution of
\begin{align} 
C'_0(\tau_0)=0.
\end{align} 
Next $\tau_1$ is given by
\begin{align} 
 \tau_1=
-\frac{ C'_2(\tau_0) }{C''_0(\tau_0)}
|u_1|^2 
-\frac{ C'_4(\tau_0) }{C''_0(\tau_0)}
|u_1|^4 .
\end{align} 
By putting this solution into the effective action (\ref{eff-action-gen}), we obtain an effective action for the Wilson loop,
\begin{align} 
S(u_1)/DN^2=C_0(\tau_0)+C_2(\tau_0) |u_1|^2
+\left(C_4(\tau_0) -\frac{1}{2} \frac{ C_2^{'2} (\tau_0)}{C''_0(\tau_0)}\right) 
|u_1|^4+\cdots.
\label{formal eff action}
\end{align} 
The sign of the coefficient of $|u_1|^4$ fixes the phase structure though the Landau-Ginzburg type argument in \cite{Aharony:2003sx, AlvarezGaume:2005fv}.
If it is positive, three phases (uniform, non-uniform and gapped) will appear and the phase transitions between them are second and third order.
If it is negative, two phases (uniform and gapped) will appear and the phase transition is first order.

In the $1/D$ expansion, since $C_4$ is very small, the sign of $C''_0(\tau_0)$ is relevant in (\ref{formal eff action}).
In our case, from (\ref{C0}), $C''_0(\tau_0)$ is a negative constant and the phase structure is fixed as the former case.

Note that $C_0(\tau)$ can be regarded as an effective action for the condensation $\triangle$ in the uniform phase.
Thus the discussion in this appendix implies that this effective action determines the phase structure.
This observation is interesting but physical meaning is unclear yet.

\section{Calculation of the $1/D$ correction}
\label{app 1/d}
In this appendix, we show the derivation of the effective action (\ref{s-eff-all}) involving the subleading $1/D$ corrections.
The analysis is similar to the one shown in appendix E of \cite{Mandal:2009vz} and we do not show the details here.

The chemical potential dependence appears only through the propagator of the complex adjoint scalars
\begin{align} 
\langle \Phi_{ij}^{\dagger I}(t) \Phi_{kl}^J(0) \rangle
 &= \frac{
  e^{ (  i    (\alpha_j-\alpha_i)+\mu)||t|| }
 }{2 \triangle}
\Biggl[
 \frac{e^{
 -  \triangle ||t|| 
  }}{1-e^{i\beta  (\alpha_j-\alpha_i)}e^{-\beta (\triangle-\mu)}}
-\frac{e^{
   \triangle ||t|| 
  }}{1-e^{i\beta  (\alpha_j-\alpha_i)}e^{\beta (\triangle+\mu)}}\Biggr]\delta_{il} 
\delta_{jk} \delta^{IJ} .
 \end{align} 
Here  $||t||$ denotes $||t+n \beta||=t$ for $0\le t < \beta$.
By using this propagator, we can evaluate the $1/D$ corrections to the effective action from loop diagrams of $\Phi^I_a$, $Y^i_a$ and $b_{ab}$.

If $k_1=0$ ($D=2\tilde{D}$), we do not need to consider $Y^i_a$ and the $(m+1)$-loop correction to the effective action is calculated as \cite{Mandal:2009vz}
\begin{align} 
-d_m\frac{(-)^m}{2m}
\left( \beta  g^2D  N\right)^m 
\sum_{n=-\infty}^\infty  \sum_{i,j=1}^N  \left( G^{(2)}_{n,ij} \right)^m, 
\label{effective action general loop}
\end{align} 
where $d_m$ is a factor derived from a property of $M_{ab,cd}$ as,
\begin{align} 
d_1=-1,~~d_2=3,~~d_m=1~~(m \ge 3) \nonumber.
\end{align} 
$G^{(2)}_{n,ik}$ is defined as
\begin{align} 
G^{(2)}_{n,ik}=\frac{1}{8\triangle^2}
\left( P^-_{n,ik}S^-_{ik} + P^+_{n,ik} S^+_{ik}+Q_{n,ik}S_{Q,ik} \right)  ,
\label{sol composite propagator}
\end{align} 
where $n$ dependent term $P^-_{n,ik}, P^+_{n,ik}$ and $Q_{n,ik}$ are defined as
\begin{align} 
P^-_{n,ik}&=\frac{1}{\pi i}\frac{-1}{\frac{i\beta 
(\alpha_k-\alpha_i)-2\triangle \beta}{2\pi i}-n },\qquad
P^+_{n,ik}=\frac{1}{\pi i}\frac{1}{\frac{i\beta 
(\alpha_k-\alpha_i)+2\triangle \beta}{2\pi i}-n },  \nonumber \\
Q_{n,ik}&=\frac{1}{\pi i}\frac{1}{\frac{i\beta (\alpha_k-\alpha_i)}{
2\pi i}-n } ,
\end{align} 
and $n$ independent term $S^-_{ik}, S^+_{ik}$ and $S_{Q,ik}$ are
\begin{align} 
 S^{+}_{il}=&1+\sum_{m=1}x^{m}\left( y^{-m}u_{-m}u_{m}^i+ 
y^{m}u_{m} u_{-m}^l \right) ,\\
 S^{-}_{il}=&1+\sum_{m=1}x^{m}\left( y^{m}u_{m}u_{-m}^i+ 
y^{-m}u_{-m} u_{m}^l \right)  ,\\
 S_{Q,il}=&x\sum_{k,m=0}x^{k+m}\left( y^{k+m+1}u_{k+m+1}u_{-k}^i u_{-m}^l
(u_{-1}^i-u_{-1}^l) 
+ y^{-k-m-1}u_{-k-m-1}u_{k}^i u_{m}^l
(u_{1}^l-u_{1}^i) \right)  ,
\end{align} 
where $x=e^{-\beta \triangle}$ and $y=e^{-\beta \mu}$.

If $k_1 \ne 0$, we have to evaluate the contribution of $Y^I_a$ also, but 
the same formula (\ref{effective action general loop}) is still available with simple changes,
\begin{align} 
S_{Q,il}& \to  
 k_1 \left.S_{Q,il}\right|_{\mu=0}+k_2 S_{Q,il} , \nonumber \\ 
S_{il}^{+}& \to  
 k_1 \left.S_{il}^{+}\right|_{\mu=0}+k_2 S_{il}^{+} , \nonumber \\ 
S_{il}^{-}& \to  
 k_1 \left.S_{il}^{-}\right|_{\mu=0}+k_2 S_{il}^{-}. \nonumber 
\end{align} 

By using technique in \cite{Mandal:2009vz}, we can evaluate (\ref{effective action general loop}).
The results in $\mu=0$ case is given by (E.30)-(E.32) in \cite{Mandal:2009vz}.
In the non-zero chemical potential case, the modifications on (E.30)-(E.32) are summarized as,
\begin{itemize}
 \item  (E.30) is not modified.
\item  (E.31) is modified as $x \to x \left(k_1+k_2 \frac{y+y^{-1}}{2}  \right) $ only.
\item (E.32) is modified as
\end{itemize}
\begin{align} 
&-(-)^n d_n N^2  \beta \triangle x^2|u_1|^4 
\frac{n-1}{2} 
\left( \frac{\tilde{\lambda} }{4 \triangle^3} \right)^n \nonumber \\
&\times \left[
\left( 
\frac{(2n-3)!!}{(2n-2)!!}+2+\triangle \beta 
 \right) \left(k_1+k_2 \frac{y+y^{-1}}{2}  \right)^2 
+\left( \frac{(2n-5)!!}{(2n-4)!!}+\triangle \beta \right) k_2^2\left(  \frac{y-y^{-1}}{2}\right)^2
\right]   
+O(x^3).
 \label{n+1 u4}
\end{align} 
Then by summing over $n$ and adding the leading results (\ref{potential triangle}), we  obtain the effective action (\ref{s-eff-all}).

\section{Expressions for $f_d(x)$}
\label{app fd}
In this section, we show the expressions for $f_d(x)$ up to $d=5$, which are important to show the existence of the condensation in the higher dimensional gauge theories.
$f_d(x)$ is given by 
\begin{align} 
f_{d}(x) \equiv& \Lambda^{3-d}  \int_0^\Lambda dp \frac{ p^{d-2}
}{ p^2+x \Lambda^2 }  .
\end{align} 
Then we can calculate it as
\begin{align} 
f_{2}(x)=& \frac{1}{\sqrt{x}} \arctan \left( 1/\sqrt{x} \right) , \nonumber \\
f_{3}(x)=& \frac{1}{2} \log \left(1+ 1/x \right) , \nonumber \\
f_{4}(x)=& 1-\sqrt{x} \arctan \left( 1/\sqrt{x} \right)  ,\nonumber \\
f_{5}(x)=& \frac{1}{2}-\frac{1}{2}x \log \left(1+ 1/x \right)    .\nonumber
\end{align} 
These satisfy the relation (\ref{fd}).


\bibliography{chemical}
\bibliographystyle{JHEP}  

\end{document}